\documentclass[twocolumn,superscriptaddress,10pt]{revtex4-1}
\usepackage{amsmath, amsfonts, amssymb, graphicx, xcolor, cancel, float, paralist,lineno}

\graphicspath{{./figures/}}

\newcommand{\Pdata}{P_{\text{data}}}
\newcommand{\Pgen}{P_{\text{gen}}}
\newcommand{\Ppost}{P_{\text{post}}}
\newcommand{\TP}{\hat{\text{TP}}}
\newcommand{\FP}{\hat{\text{FP}}}
\newcommand{\TN}{\hat{\text{TN}}}
\newcommand{\FN}{\hat{\text{FN}}}
\newcommand{\VJl}{VJ$l$ }

\newcommand{\equal}{These authors contributed equally.}
\newcommand{\TM}{}
\newcommand{\GA}{}

\usepackage{ifthen}
\newboolean{elife}
\setboolean{elife}{false}

\newcommand{\SIFIG}[2]{Fig.~\ref{figsupp:#2}}
\newcommand{\FIG}[1]{Fig.~\ref{fig:#1}}
\newcommand{\METH}[1]{Methods section~\ref{#1}}

\begin{document}

\author{Natanael~Spisak}
\thanks{
  Current address: 
  Department of Biological Sciences, 
  Columbia \mbox{University}, New York, New York, USA.
}
\affiliation{
  Laboratoire de Physique de l'\'Ecole normale sup\'erieure, CNRS, PSL University,\\
  Sorbonne Universit\'e, and Universit\'e Paris Cit\'e, Paris, France}
\author{Gabriel~Ath\`enes}
\affiliation{
  Laboratoire de Physique de l'\'Ecole normale sup\'erieure, CNRS, PSL University,\\
  Sorbonne Universit\'e, and Universit\'e Paris Cit\'e, Paris, France}
\affiliation{Saber Bio SAS, Institut du Cerveau,
	iPEPS The HealthtechHub,
	Paris, France}
\author{Thomas~Dupic}
\affiliation{
  Department of Organismic and Evolutionary Biology,
  Harvard University, Cambridge, United States}
\author{Thierry~Mora}
\thanks{\equal}
\affiliation{
  Laboratoire de Physique de l'\'Ecole normale sup\'erieure, CNRS, PSL University,\\
  Sorbonne Universit\'e, and Universit\'e Paris Cit\'e, Paris, France}
\author{Aleksandra~M.~Walczak}
\thanks{\equal}
\affiliation{
  Laboratoire de Physique de l'\'Ecole normale sup\'erieure, CNRS, PSL University,\\
  Sorbonne Universit\'e, and Universit\'e Paris Cit\'e, Paris, France}

\newcommand{\deftitle}{{Combining mutation and recombination statistics to infer \\ clonal families in antibody repertoires}}

\title{\deftitle}

\begin{abstract}
B-cell repertoires are characterized by a diverse set of receptors of
distinct specificities generated through two
processes of somatic diversification: V(D)J recombination and somatic
hypermutations. B cell clonal families 
stem from the same V(D)J recombination event, but 
differ in their hypermutations. 
Clonal families identification is key to understanding  B-cell repertoire function, evolution and
dynamics. 
We present HILARy (High-precision Inference of Lineages
in Antibody Repertoires), an efficient, fast and precise method to identify clonal families from
{\TM single- or paired-chain repertoire} sequencing datasets.
HILARy combines probabilistic models that capture the receptor generation and selection statistics   with adapted clustering methods
to achieve consistently high inference accuracy. 
 It automatically leverages
the phylogenetic signal of shared mutations in difficult repertoire subsets. 
 Exploiting the high sensitivity of the method, we find the
statistics of evolutionary properties such as the site frequency
spectrum and $d_N/d_S$ ratio do not depend on the junction length. We also identify a broad range of selection pressures spanning two orders of magnitude.

\end{abstract}

\maketitle
\section{Introduction}
B cells play a key role in the adaptive immune response through their diverse repertoire of immunoglobulins (Ig). These proteins recognize foreign pathogens in their membrane-bound form (called B-cell receptor or BCR), and battle them in their soluble form (antibody). Each B cell expresses a unique BCR that can bind their antigenic targets with high affinity. The set of distinct BCR harbored by the organism is highly diverse \citep{Briney2019}, thanks to two processes of diversification: V(D)J recombination and somatic hypermutation. These stochastic processes ensure that repertoires can match a variety of potential threats, including proteins of bacterial and viral origin that have never been encountered before.

\begin{figure}
\includegraphics[width=.5\textwidth]{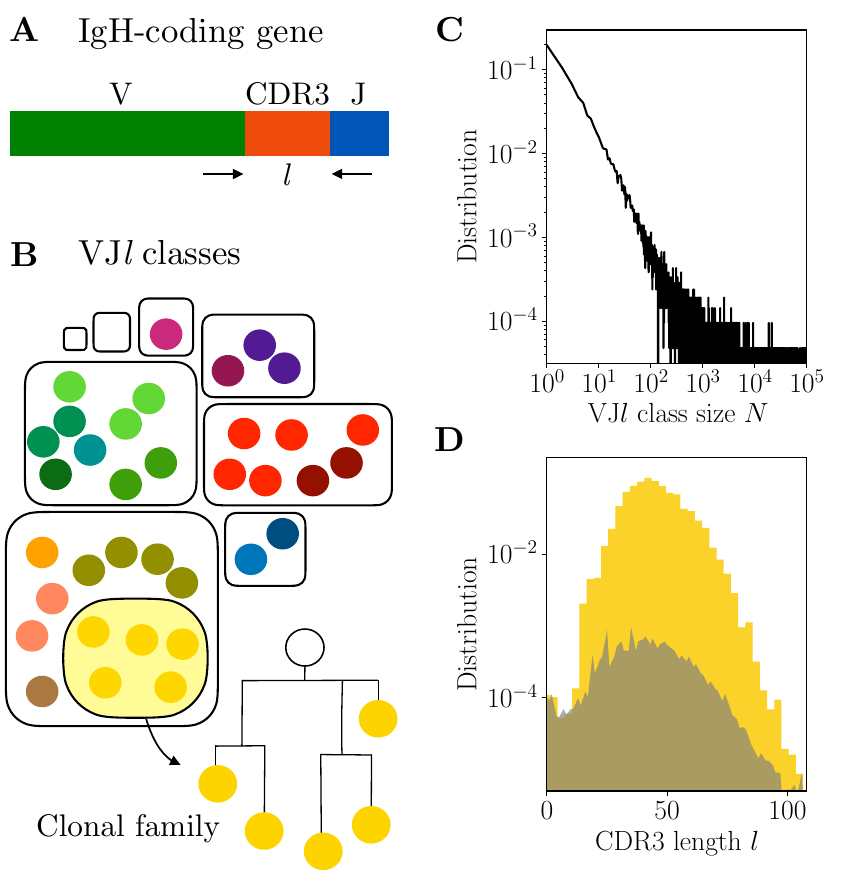}
\caption{
{\bf Clonal families and \VJl classes.}
(A) Variable region of the IgH-coding gene. 
(B) A clonal family is a lineage of related B cells stemming from the same VDJ recombination event. The partition of the BCR repertoire into clonal families is a refinement of the partition into \VJl classes, defined by sequences with the same V and J usage and the same CDR3 length $l$.
(C-D) Properties of \VJl classes in donor 326651 from \cite{Briney2019}.
(C) Distribution of \VJl class sizes exhibits power-law scaling. The total number of pairwise comparisons in the largest \VJl classes is $\sim {10^5}^2= 10^{10}$.
(D) Distribution of the CDR3 length $l$. The distribution is in yellow for in-frame CDR3 sequences ($l$ multiple of 3), and in gray for out-of-frame sequences. 
}
\label{fig:figure1}  
\end{figure}

V(D)J recombination takes place during B cell differentiation \citep{Hozumi1976, Schatz2011}. For each Ig chain, V, D, and J gene segments for the heavy chain, and V and J gene segments for the light chain, are randomly chosen and joined with random non-templated deletions and insertions at the junction, creating a long, hypervariable region, called the Complementarity Determining Region 3 (CDR3) (\FIG{figure1}A). Cells are subsequently selected for the binding properties of their receptors and against autoreactivity.
At this stage, the repertoire already covers a wide range of specificities. In response to antigenic stimuli, B cells with the relevant specificities are recruited to germinal centers, where they proliferate and their Ig-coding genes undergo somatic hypermutation \citep{victora2022germinal} in the process of affinity maturation. Somatic hypermutation consists primarily of point substitutions, as well as insertions and deletions, restricted to Ig-coding genes \citep{Feng2020}. The mutants are selected for high affinity to the particular antigenic target, and the best binders further differentiate into plasma cells and produce high-affinity antibodies. A more diverse pool of variants forms the memory repertoire, leaving an imprint of the immune response that can be recalled upon repeated stimulation.

A clonal family is defined as a collection of cells that stem from a unique V(D)J rearrangement, and has diversified as a result of hypermutation, forming a lineage (\FIG{figure1}B). These families are the main building blocks of the repertoire. Since members of the same family usually share their specificities \citep{DeBoer2001}, affinity maturation first competes families against each other for antigen binding in the early stages of the reaction, and then selects out the best binders within families in the later stages \citep{tas2016visualizing, Mesin2016}.

High-throughput sequencing of single receptor chains offers unprecedented insight into the diversity and dynamics of the repertoire. Recent experiments have sampled the repertoires of the immunoglobulin heavy chain (IgH) of healthy individuals at great depth to reveal their structure \citep{Briney2019}. Disease-specific cohorts are now routinely subject to repertoire sequencing studies, which help to quantify and understand the dynamics of the B-cell response \citep{kreer2020longitudinal, nielsen2020human}.

Partitioning BCR repertoire sequence datasets into clonal families is a critical step in understanding the architecture of each sample and interpreting the results. Identifying these lineages allows for quantifying selection \citep{Yaari2012a, Yaari2015, ortega2021modeling} and for detecting changes in longitudinal measurements \citep{nielsen2020human, turner2020human}. 
In recent years, many strategies have been developed that take advantage of CDR3 hypervariability \citep{abdollahi2020automatic}: it is generally unlikely that the same or a similar CDR3 sequence be generated independently multiple times \citep{Elhanati2015, ortega2021modeling}. Other approaches make use of the information encoded in the intra-lineage patterns of divergence due to mutations \citep{briney2016clonify, nouri2020somatic}. All inference techniques need to balance accuracy and speed. Simpler methods are fast but have low precision {\TM (also called positive predictive value)} while more complex algorithms have long computation times that do not scale well with the number of sequences. This prohibits the analysis of recent large-scale data such as \cite{Briney2019}.

In this work, we propose a new method for inferring clonal families from high-throughput sequencing data that is both fast and accurate. We use probabilistic models of junctional diversity to estimate the level of clonality in repertoire subsets, allowing us to tune the sensitivity threshold {\em a priori} to achieve a desired accuracy.
We have developed two complementary algorithms. The first one {\TM (HILARy-CDR3)} uses a very fast CDR3-based approach that avoids pairwise comparisons, while the second one {\TM (HILARy-full)} additionally exploits information encoded in the phylogenetic signal outside of the junction. We compare our method with state-of-the-art approaches in~a~benchmark with realistic synthetic~data.

\section{Results}

\subsection{Analysis of pairwise distances within \VJl classes} 

A common strategy for partitioning a BCR repertoire dataset into clonal families is to go through all pairs of sequences and identify pairs of clonally related sequences. In the following, we call such related pairs {\em positive}, and pairs of sequences belonging to different families {\em negative}. Then, the partition is built by single-linkage clustering, {\GA which consists of} recursively grouping all positive pairs. Two characteristics of the repertoire complicate the search for this partition: large total number of pairs, and low proportion of positive pairs. 
{\TM In this section we analyze and model the statistics of pairs of sequences in natural repertoires to inform our choice of the clustering method and parameters. In the next section we will leverage that analysis to design an optimized clustering procedure. To help following notations, a summary of their definitions is provided in Table~\ref{tab}.}

\begin{table}
{\TM 
  \begin{tabular}{c|c}
    & {\bf definition}\\
    \hline\hline
    $\rho$ & prevalence/fraction of positive pairs\\
    \hline
    $\pi$ & precision = TP / (TP + FP)\\
    \hline
    $s$ & sensitivity = TP / (TP + FN)\\
    \hline
    $p$ & fallout = FP / (FP + TN)\\
    \hline
    $t$ & threshold on CDR3 distance\\
    \hline
    $l$ & CDR3 length\\
    \hline
    $n$ & CDR3 Hamming distance of a pair\\
    \hline
    $x$ & normalized CDR3 Hamming distance $=l/n$\\
    \hline
    $x'$ & CDR3 Hamming distance, centered and scaled \\
    \hline
    $y'$ & shared mutations on V segment, centered and scaled \\
    \hline
    $\mu$ & mean $x$ between positive pairs\\
    \hline
    $P_{\rm T}$ & model distribution for positive pairs\\
    \hline
    $P_{\rm F}$ & model distribution for negative pairs
\end{tabular}  
\caption{\TM Summary of notations used throughout the paper. Hats ${\hat{\ }}$denote estimates from the fit of the mixture model. Stars $^*$ denote estimates after imposing $99\%$ precision. The ``post'' subscript denotes quantities after applying single-linkage clustering to obtain a partition from positive pairs.\label{tab}}
}
\end{table}

\newcommand{\figtwocaption}{
{\bf CDR3-based inference method {\TM (HILARy-CDR3)}}. 
(A)~Example distribution of normalized Hamming distances, $x=n/l$, for one \VJl class with CDR3 length $l=21$, V gene IGHV3-9 and J gene IGHJ4 (black). We fit the distribution by a mixture of positive pairs (belonging to the same family, in blue), and negative pairs (belonging to different families, in red). See \SIFIG{figure2}{figure5} for example fit results across different CDR3 lengths.
Inset:  the prevalence is defined as a fraction of positive pairs and was estimated to $\hat{\rho}=3.1\%$. Data from donor 326651 of \cite{Briney2019}.
(B)~Distribution of the maximum likelihood estimates of prevalence $\hat{\rho}$ across \VJl classes {\TM in donor 326651}.
(C-F)~The choice of threshold $t$ on the normalized Hamming distance $x$ translates to the following {\em a priori} characteristics of inference {(\TM illustrated here for arbitrarily chosen $\rho$ and $\mu$)}.
(C)~Fallout rate $\hat{p}(t)=\hat{\text{FP}}/(\hat{\text{FP}}+\hat{\text{TN}})$. The null distribution of all negatives (N=FP+TN) is estimated using the soNNia sequence generation software. 
(D)~Sensitivity $\hat{s}(t)=\hat{\text{TP}}/(\hat{\text{TP}}+\hat{\text{FN}})$.  
(E-F)~Precision $\hat{\pi}=\hat{\text{TP}}/(\hat{\text{TP}}+\hat{\text{FP}})$. For the same choice of threshold $t$, a low prevalence of $\hat{\rho}=10^{-3}$~(E) leads to lower precision than high prevalence of $\hat{\rho}=10^{-1}$~(F).
(G)~{\TM Model} distribution $P_{\rm F}(x|l)$ of distances between unrelated sequences, for $l=15,30,45,60$, computed by the soNNia software.
(H)~{\TM Precision $\hat{\pi}$, computed {\em a priori} (i.e. before doing the inference) from the model with $\hat\mu=0.04$, $\hat\rho=0.1$, and $l=15,30,45,60$ (colors as in G), as a function of the threshold $t$. For each \VJl class and its own inferred $\hat\rho$ and $\hat\mu$, the threshold $t$ is chosen to achieve a desired $\pi^*$.}
(I)~High-precision threshold $t^*$ ensuring $\hat{\pi}(t^*)=\pi^*=99\%$ {\TM \em a priori}, as a function of CDR3 length $l$ for different values of the prevalence $\hat{\rho}$, and $\hat\mu=0.04$, {\TM as predicted by the model}.
(J)~Sensitivity $\hat{s}(t^*)$ at the high-precision threshold $t^*$, as a function of CDR3 length $l$ for different values of the prevalence $\hat{\rho}$ (colors as in I). Solid lines denote {\TM\em a priori} prediction for intermediate mean distance $\mu=4\%$, dashed lines denote {\TM actual performance of HILARy-CDR3} in a synthetic dataset.}

\newcommand{\figtwoSonetitle}{Mean intra-family distances.}
\newcommand{\figtwoSonecaption}{Distribution of the maximum likelihood estimates of mean intra-family distance $\hat{\mu}$ across \VJl classes.}

\newcommand{\figtwoStwotitle}{Null distribution $P_{\rm N}(x|l)$}
\newcommand{\figtwoStwocaption}{of CDR3 distances between unrelated sequences for $l\in[15,81]$, computed by soNNia software. White line denotes a growing threshold ensuring a fallout rate $p<10^{-4}$ as determined by this distribution.}

\newcommand{\figtwoSthreetitle}{Distribution of normalized Hamming distances}
\newcommand{\figtwoSthreecaption}{$x=n/l$, for largest \VJl class for each CDR3 length $l=15,\dots,81$ (black). We fit the distribution by a mixture of positive pairs ($P_{\rm T}(x|\mu)$ in blue), and negative pairs ($P_{\rm N}(x)$, in red). For $l=18$ the estimate $\hat{\mu}$ is too large results in large fitting error and for sensitivity computation we used global $\mu=4\%$ (in green).}

\newcommand{\figtwoSfourtitle}{Distribution of post-selection probabilities}
\newcommand{\figtwoSfourcaption}{$\Ppost$ of CDR3 nucleotide sequences computed using soNNia across CDR3 lengths. Short junctions are on average more likely to be generated in VDJ recombination and pass subsequent selection \citep{Isacchini2021}. This makes inference in low-$l$ classes more difficult, a feature reflected by synthetic dataset constructed by sampling unmutated lineage progenitors from the soNNia model.}

\newcommand{\figtwoSfivetitle}{Site frequency spectra}
\newcommand{\figtwoSfivecaption}{estimated for families identified
  using high-precision CDR3-based inference method (HILARy-CDR3) in the subset of the data where this approach is highly reliable (large-$l$ and large-$\hat{\rho}$ regime). The distributions are shown for families of varying family size, $z\in[10,100]$ and averaged over all families of a given size. Together with the exact configuration of sequences carrying a given substitution, synthetic datasets of the same signatures of mutations and clonal expansions can be generated.}

\newcommand{\figtwoSsixtitle}{Distribution of normalized Hamming distances}
\newcommand{\figtwoSsixcaption}{$x=n/l$, for $l$ classes, averaging over all \VJl classes. We fit the distribution by a mixture of positive pairs using a geometric distribution ($P_{\rm T}(x|\mu)$ in blue), and negative pairs ($P_{\rm N}(x)$, in red). The corresponding prevalence estimates $\hat{\rho}$ are used for small \VJl classes for which this parameter cannot be reliably estimated independently.}

\newcommand{\figtwoSseventitle}{Prevalence and \VJl class.}
\newcommand{\figtwoSsevencaption}{Dependence of prevalence estimates $\hat{\rho}$ on \VJl class size $N$ for largest classes in donor 326651 from \cite{Briney2019}. 28\% of variation in prevalence estimates can be explained by variation in \VJl class sizes.}

\ifthenelse{\boolean{elife}}{
\begin{figure}
\includegraphics[width=\linewidth]{figure2.pdf}
\caption{\figtwocaption}
\label{fig:figure2}
\figsupp[{\figtwoSonetitle}]
{{\bf \figtwoSonetitle} \figtwoSonecaption}
{\includegraphics[width=0.5\linewidth]{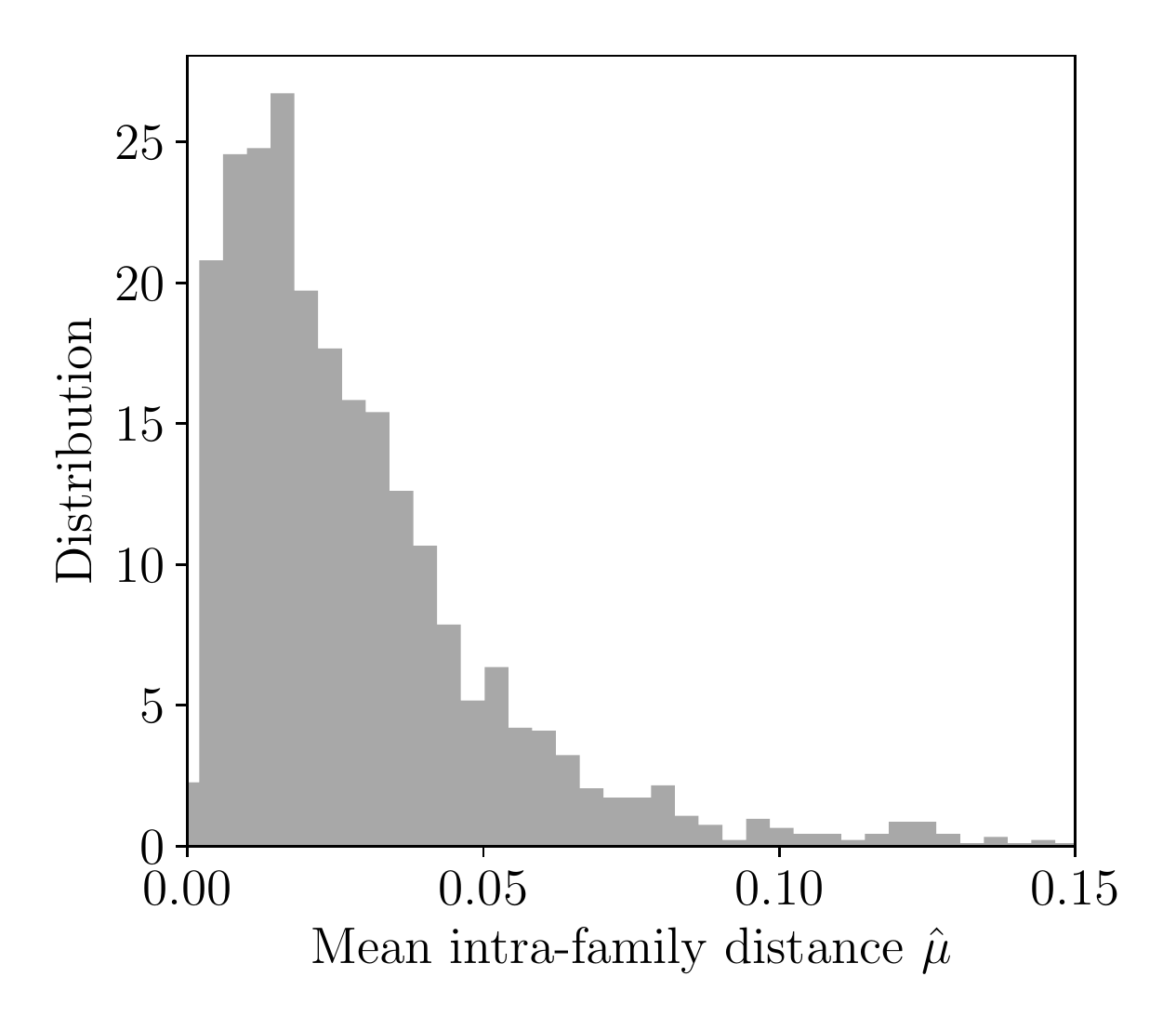}}
\label{figsupp:figure1}
\figsupp[{\figtwoStwotitle.}]
{{\bf \figtwoStwotitle} \figtwoStwocaption}
{\includegraphics[width=0.5\linewidth]{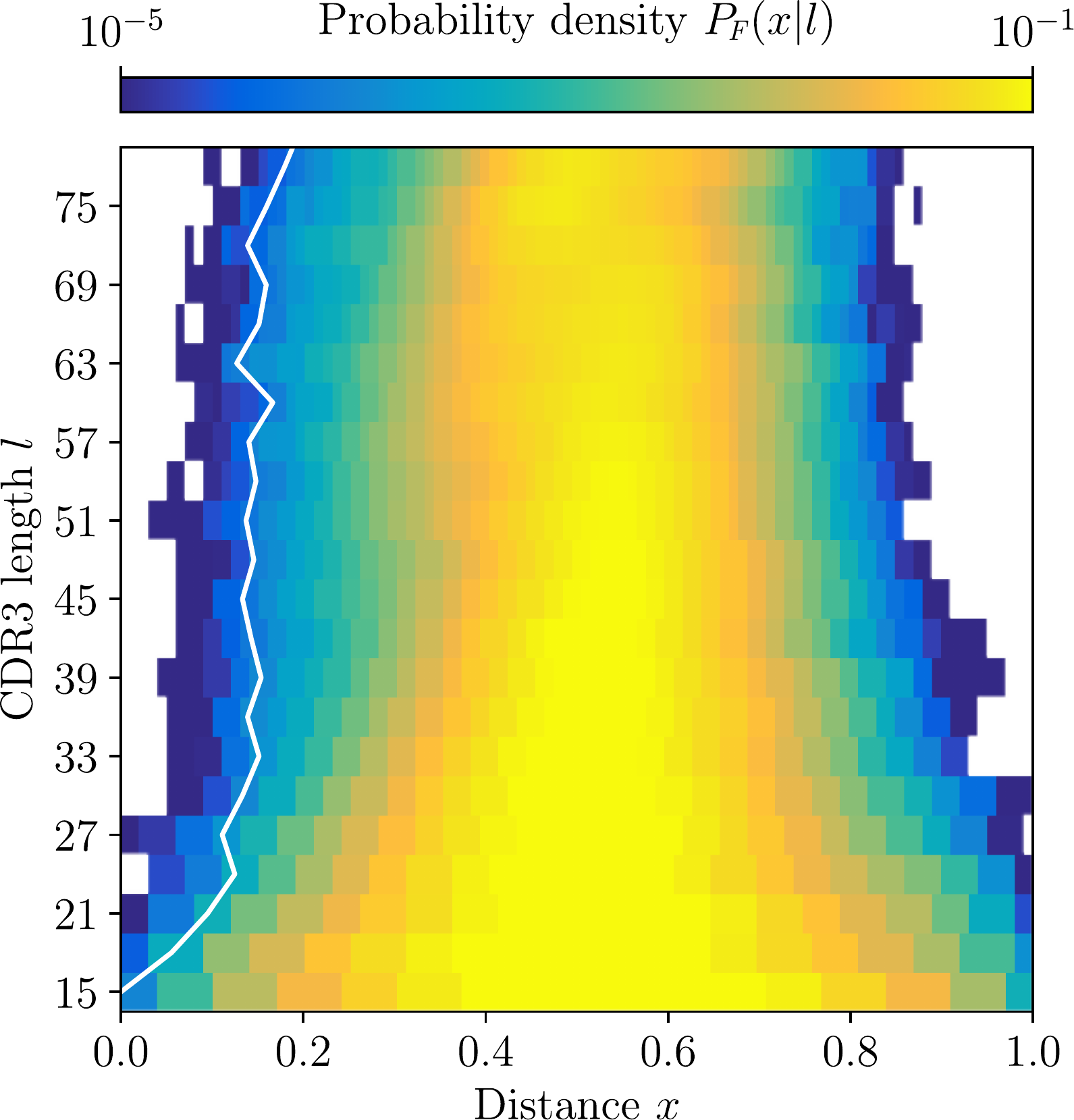}}
\label{figsupp:figure2}
\figsupp[{\figtwoSthreetitle.}]
{{\bf \figtwoSthreetitle} \figtwoSthreecaption}
{\includegraphics[width=\linewidth]{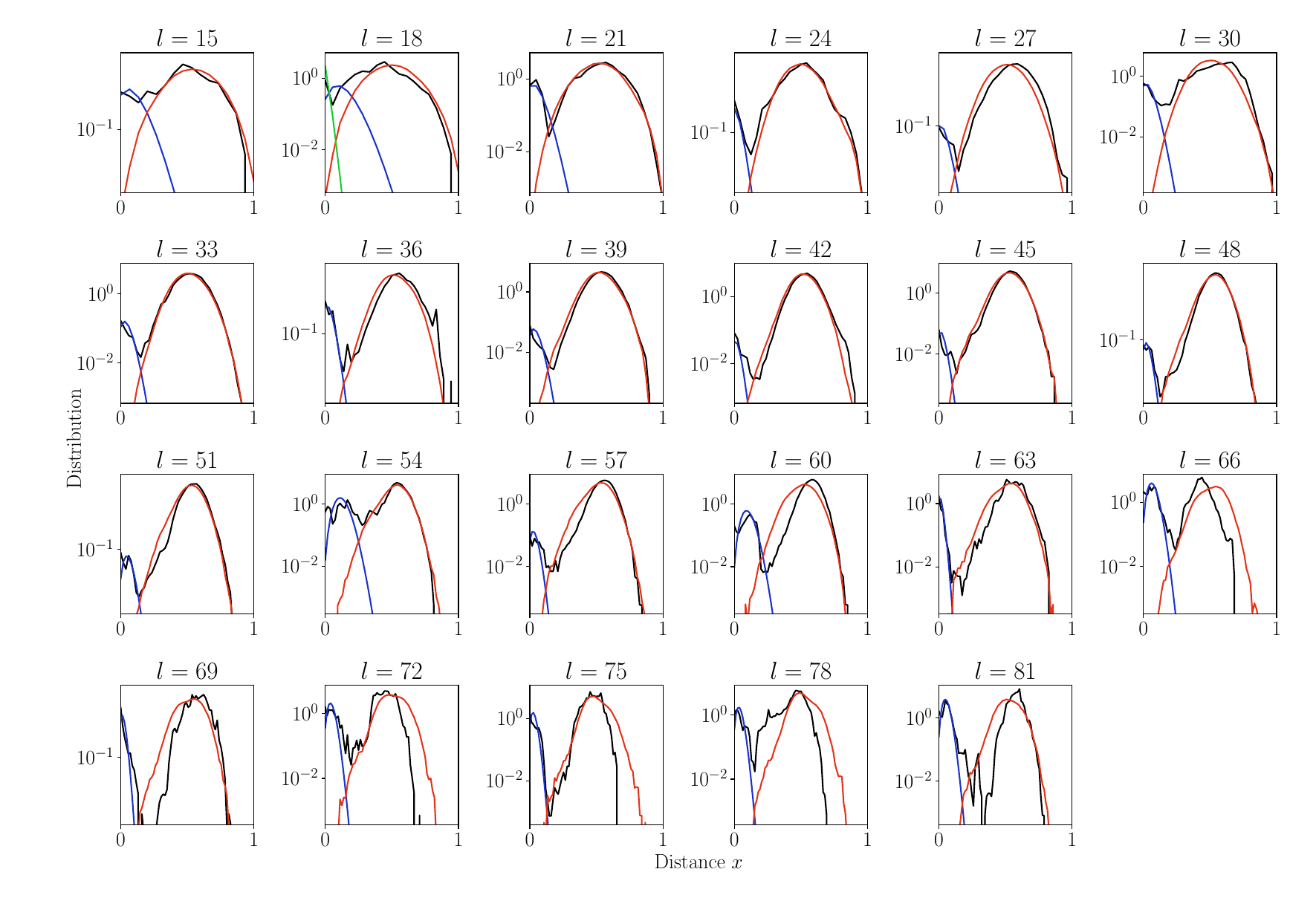}}
\label{figsupp:figure3}
\figsupp[{\figtwoSfourtitle.}]
{{\bf \figtwoSfourtitle} \figtwoSfourcaption}
{\includegraphics[width=0.5\linewidth]{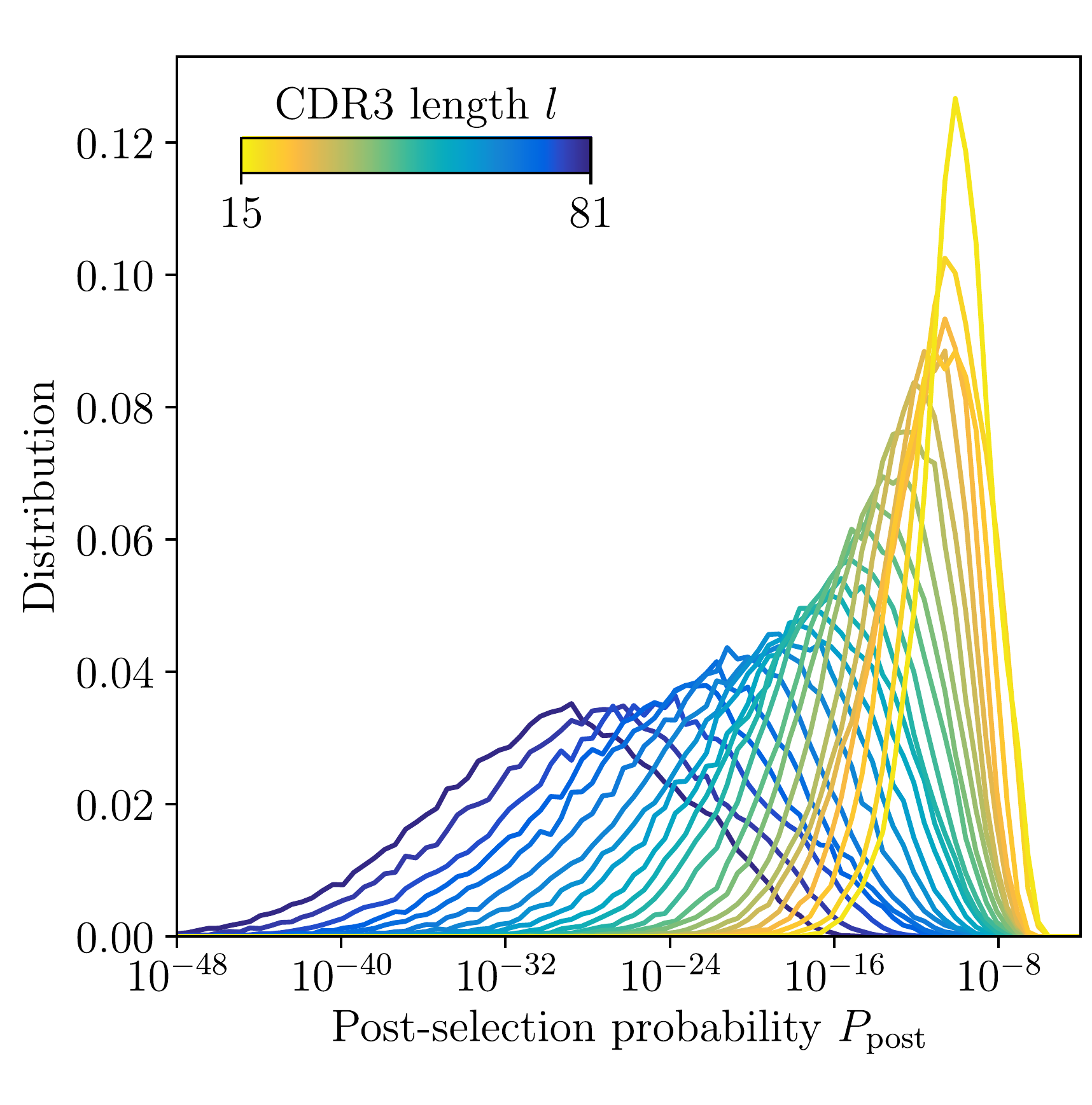}}
\label{figsupp:figure4}
\figsupp[{\figtwoSfivetitle.}]
{{\bf\figtwoSfivetitle} \figtwoSfivecaption}
{\includegraphics[width=0.5\linewidth]{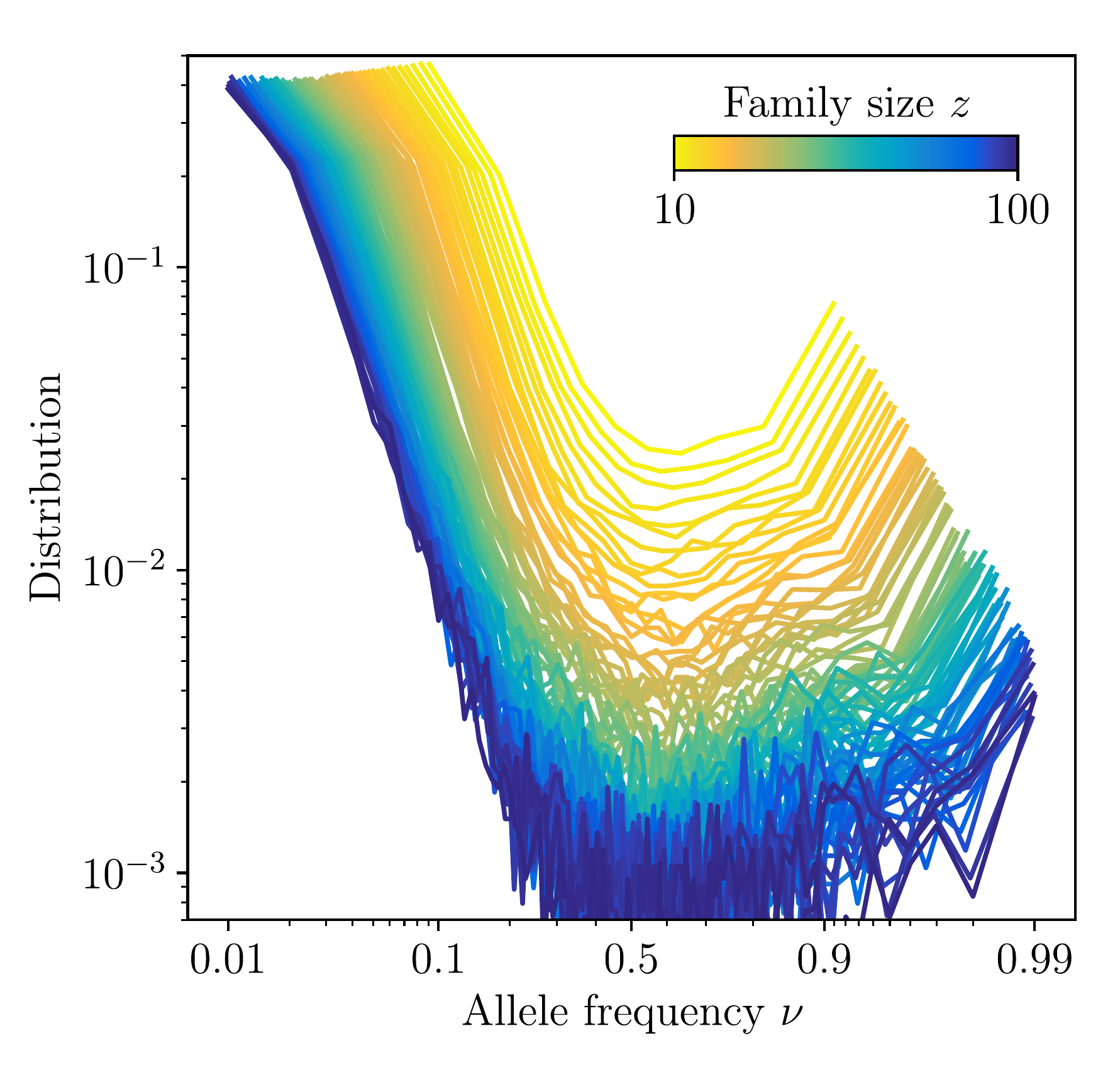}}
\label{figsupp:figure5}
\figsupp[{\figtwoSsixtitle.}]
{{\bf\figtwoSsixtitle} \figtwoSsixcaption}
{\includegraphics[width=\linewidth]{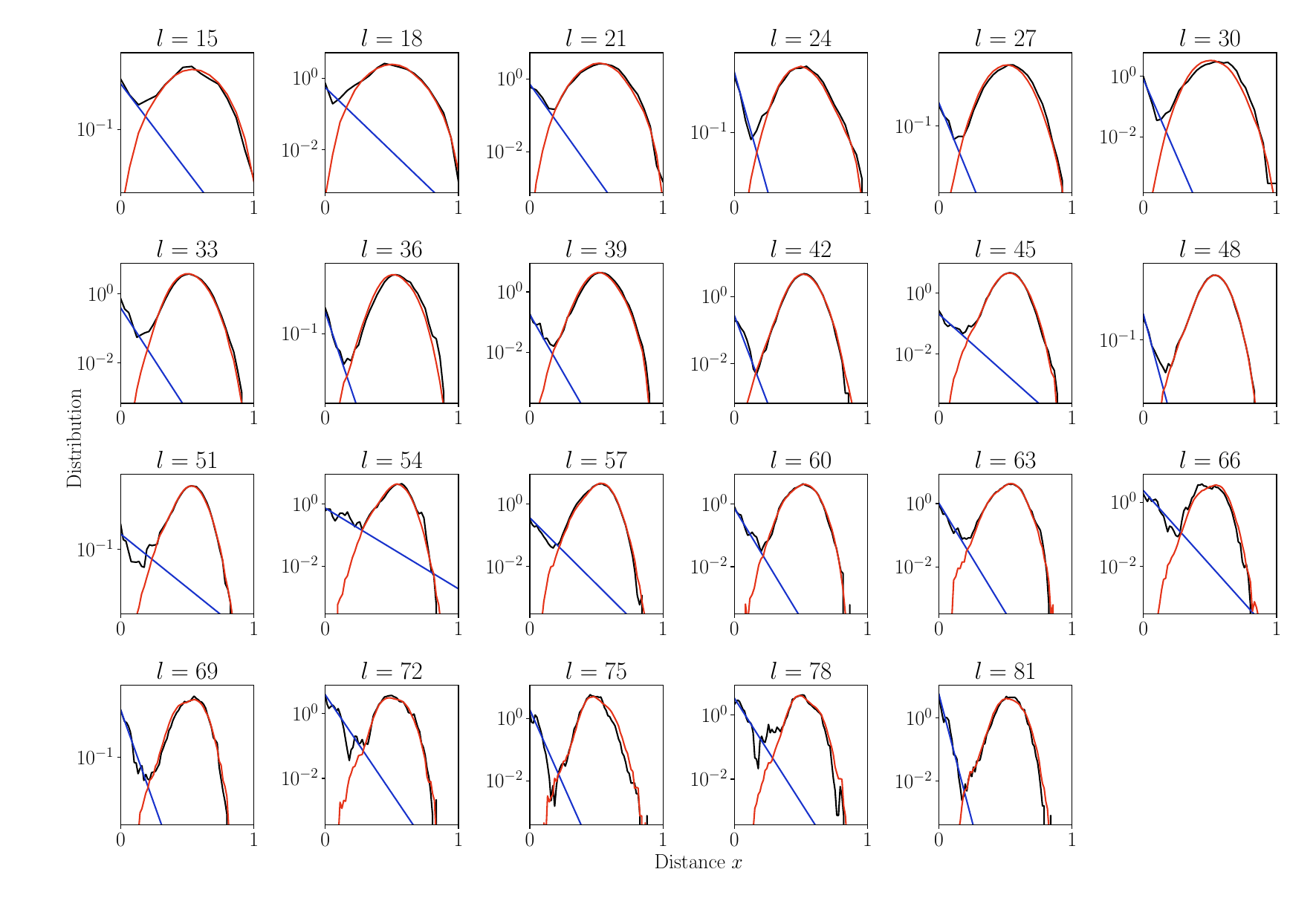}}
\label{figsupp:figure6}
\figsupp[{\figtwoSseventitle.}]
{{\bf\figtwoSseventitle} \figtwoSsevencaption}
{\includegraphics[width=0.5\linewidth]{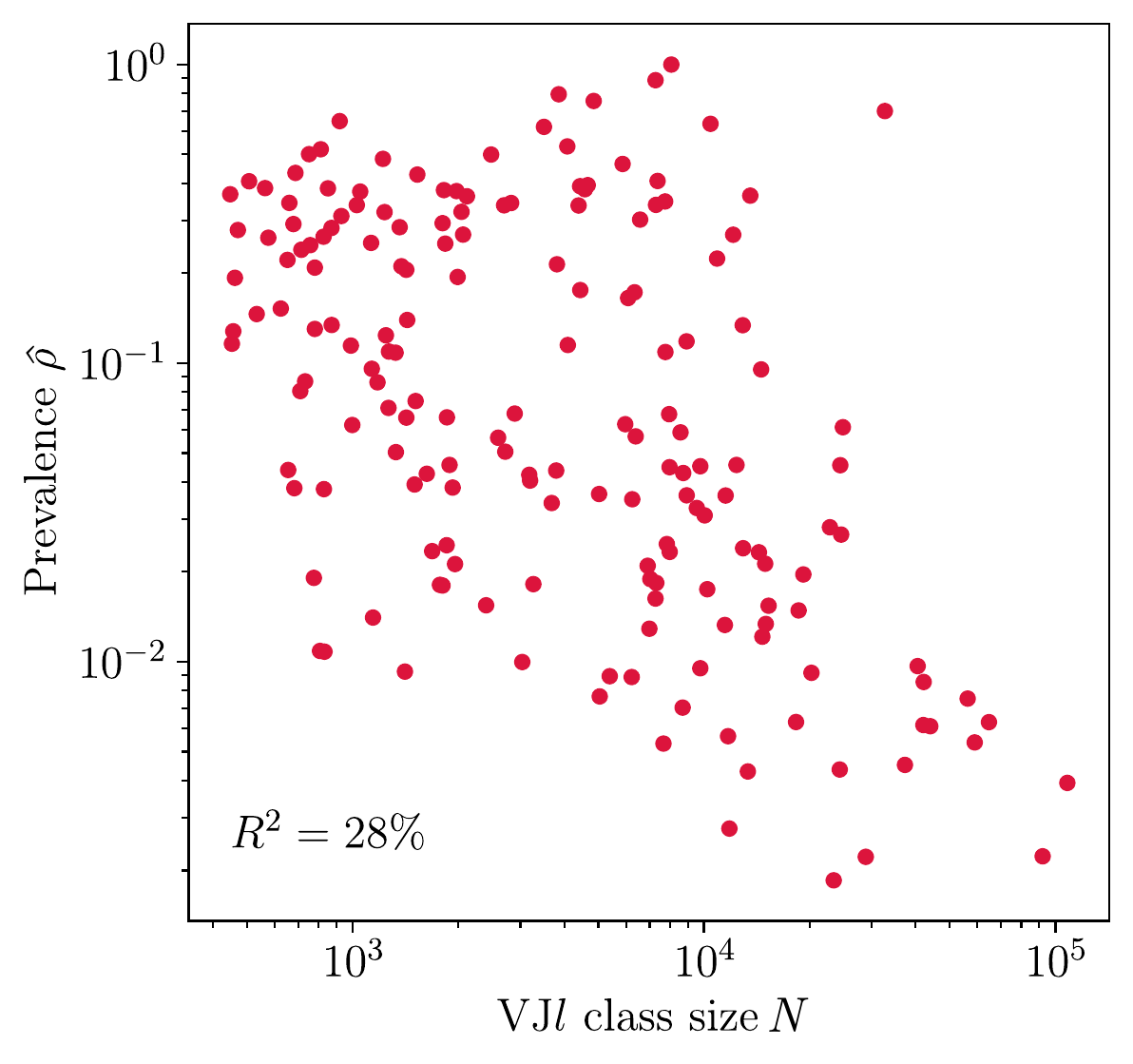}}
\label{figsupp:figure7}
\end{figure}
}
{
\begin{figure*}
\includegraphics[width=\linewidth]{figure2.pdf}
\caption{\figtwocaption}
\label{fig:figure2}
\end{figure*}
}

A pair of related sequences is expected to {\GA share the same V and J genes}, as well as the same CDR3 length $l$, as determined by alignment to the templates (\FIG{figure1}A).
The methods developed here begin by partitioning the data into \VJl classes, defined as subsets of sequences with the same V and J gene usage, and CDR3 length $l$ (\FIG{figure1}B). 
For a description of the data preprocessing and alignment to the V and J gene templates see \METH{Methods-alignment}. {\TM Clustering will then be performed within each {\TM \VJl} class independently.
  While this first step} severely limits the number of unnecessary comparisons, some \VJl classes still exceed $10^5$ sequences in large datasets, leading to the order of $10^{10}$ pairs (see \FIG{figure1}C for the distribution of the \VJl class sizes $N$ for donor 326651 of \cite{Briney2019}).

{\GA The CDR3 plays an important role in encoding the signature of the VDJ rearrangement.} As we will see, the CDR3 length $l$ has a strong impact on the difficulty of clonal family reconstruction. 
The distribution of CDR3 lengths $l$ observed in the data is shown in \FIG{figure1}D.
{\TM In what follows we restrict our analysis to sequences with CDR3 lengths a multiple of 3 and between 15 and 105, relying on the common approximation that sequences with no frameshift in the CDR3 come from a productive naive ancestor. The number of sequences with length larger than 105 is too small to reach meaningful conclusions, and sequences of length smaller than 15 are likely nonfunctional (as evidenced by the similar number of in-frame and out-of-frame sequences in \FIG{figure1}D).}

{\TM In each \VJl class, we call {\em prevalence} and denote by $\rho$ the proportion of positive pairs, i.e. the number of positive pairs divided by the total number of pairs. This quantity is unknown in the absence of the ground-truth partition. However, we can estimate it from the statistics of pairwise distances.}
We compute the Hamming distance $n$ of each pair of CDR3s, defined as the number of positions at which the two nucleotide sequences differ. The distribution of these distances normalized by the CDR3 length, denoted by $x$, shows a clear bimodal structure {\TM in data (donor 326651 of \cite{Briney2019})}, with two identifiable components (\FIG{figure2}A): the contribution of positive pairs (of proportion $\rho$) peaks near $x=0$ and decays quickly, whereas the bell-shaped contribution of negative pairs (of proportion $1-\rho$) peaks around $x = 1/2$. 

The prevalence $\rho$ can be formally written as $[\sum_i z_i(z_i-1)/2]/[N(N-1)/2]$, where $z_i$ denote the sizes of the clonal families in the \VJl class, but we do not know these sizes before the partition into families is found.
To overcome this issue, we developed a method to estimate $\rho$ {\em a priori}, without knowing the family structure (\METH{Methods-prevalence}). We do this by fitting the {\TM empirical} distribution of $x$ {\TM as a mixture model}, $P(x)=\hat{\rho} P_{\rm T}(x) + (1-\hat{\rho}) P_{\rm F}(x)$, where $P_{\rm T}(x)$ and $P_{\rm F}(x)$ are the distributions of distances between positive {\TM (T as true)} and negative {\TM (F as false)} pairs (\FIG{figure2}C and D), estimated as follows. $P_{\rm F}(x) = P_{\rm F}(x|l)$ is computed for each length $l$ by generating a large number of unrelated, same-length sequences with the soNNia model of recombination and selection \citep{Isacchini2021}, and calculating the distribution of their pairwise distances (\METH{Methods-Ppost}). $P_{\rm T}(x)$ is approximated by a Poisson distribution, $P_{\rm T}(x) = (\mu l)^{xl}e^{-\mu l}/(xl)!$, with adjustable parameter $\mu$, {\TM which is proportional to the average hypermutation rate within the clone}.
{\TM The fit of $P(x)$ by the mixture model} is performed  {\TM for each \VJl class} with an expectation-maximization algorithm which finds maximum-likelihood estimates of the prevalence $\hat{\rho}$ and mean intra-family distance $\hat{\mu}$, {\TM the only free parameters of the mixture model}.

The results of the fit to {\TM real} data {\TM (donor 326651 of \cite{Briney2019})} show that $\hat{\mu}$ varies little between \VJl classes, around {\GA$\hat\mu\simeq 4\%$} (\SIFIG{figure2}{figure1}). In contrast, the prevalence $\hat \rho$ varies widely across classes, spanning three orders of magnitude (\FIG{figure2}B). In addition, when we examine the \VJl classes with increasing CDR3 length $l$, we find {\GA that the part of the {\TM model} distribution corresponding to positive pairs, $P_{\rm T}(x)$, varies little}, whereas the {\TM model} distribution over negative pairs $P_{\rm F}(x)$ becomes more and more peaked around $1/2$ (\FIG{figure2}G and \SIFIG{figure2}{figure2}), making the two {\TM categories} more easily separable.

\subsection{CDR3-based inference method with adaptive threshold} 

We want to build a classifier between positive and negative pairs using the normalized distance $x$ alone, by setting a threshold $t$ so that pairs are called positive if $x\le t$, and negative otherwise. Using our model for $P(x)$, for any given $t$ we can evaluate the number of true positives ($\TP$) and false negatives ($\FN$) among all positive pairs ($\hat{\text{P}}=\TP+\FN$), as well as true negatives ($\TN$) and false positives ($\FP$) among the negative pairs ($\hat{\text{N}}=\TN+\FP$), as schematized in \SIFIG{figure2}{figure2}C and D.

Our goal is to set a threshold $t$ that ensures a high precision, $\hat{\pi}(t)$, defined as a proportion of true positives among all pairs classified as positive (\FIG{figure2}E). 
In a single-linkage clustering approach, we will join two clusters with at least one pair of positive sequences between them. Therefore, it is critical to limit the number of false positives, which can cause the erroneous merger of large clusters.
We can write:
\begin{equation}
\hat{\pi}(t) \equiv \frac{\hat{\text{TP}}}{\hat{\text{TP}}+\hat{\text{FP}}} = \frac{\hat{\rho} \hat{s}(t)}{\hat{\rho} \hat{s}(t) + (1-\hat{\rho}) \hat{p}(t)},
\label{precision}
\end{equation}
where $\hat{p}(t)$ is the estimate of the fall-out rate (\FIG{figure2}C), evaluated {\GA from the} soNNia-computed $P_{\rm F}$:
\begin{equation}
\hat{p}(t) \equiv \frac{\hat{\text{FP}}}{\hat{\text{N}}}=\sum_{x\le t} P_{\rm F}(x),
\label{fallout}
\end{equation}
and $\hat{s}(t)$ is the estimated sensitivity (\FIG{figure2}D), evaluated from the Poisson fit to $P_{\rm T}$ (\METH{Methods-CDR3}): 
\begin{equation}
\hat{s}(t) \equiv \frac{\hat{\text{TP}}}{\hat{\text{P}}}=\sum_{x\le t} P_{\rm T}(x).
\label{sensitivity}
\end{equation}
Finally, the estimated prevalence $\hat{\rho}\equiv{\hat{\text{P}}}/{(\hat{\text{P}}+\hat{\text{N}})}$ is inferred from the $P(x)$ distribution as explained above.

\FIG{figure2}H shows $\hat{\pi}(t)$ as a function of $t$ for different CDR3 lengths and a fixed value of $\hat \rho$. For each \VJl class, we define the threshold $t=t^*$ that reaches 99\% precision, $\hat\pi(t^*)=\pi^*=99\%$, by inverting \eqref{precision}. This adaptive threshold depends on the \VJl class through the CDR3 length $l$ and the prevalence $\rho$, and it increases with both (\FIG{figure2}I): low clonality (small $\rho$) means few positive pairs and a smaller adaptive threshold, while short CDR3 means less information and a stricter inclusion criterion.

The {\TM predicted} sensitivity, $\hat s(t^*)$, which tells us how much of the positives we are capturing, is shown in \FIG{figure2}J. We conclude that for a wide range of parameters, the method is predicted to achieve both high precision and high sensitivity. However, it is expected to fail when the prevalence and the CDR3 length are both low. At the extreme, for small values $\rho$ and $l$, even joining together identical CDR3s ($t=0$) results in poor precision because of convergent recombination (reflected by $t^*<0$).

{\TM The resulting procedure, which we call HILARy-CDR3, can be applied to Ig repertoire data through the following steps: (1) group sequences by \VJl class; (2) in each class, fit the mixture model to the distribution of pairwise distance to infer $\hat \rho$ and $\hat \mu$; (3) invert Eqs.~\ref{precision}-\ref{sensitivity} to find the high-precision threshold $t^*$; (4) classify positive and negative pairs according to that threshold; (5) complete the paritition by applying single-linkage clustering to positive pairs.}

\newcommand{\figthreetitle}{Full inference method with mutational information {\TM (HILARy-full)}.}
\newcommand{\figthreecaption}{(A)~For a pair of sequences, $n_1,n_2$ denote the numbers of mutations along the templated region (V and J), and $n_0$ is the number of shared mutations. For related sequences, $n_0$ corresponds to mutations on the initial branch of the tree, and is expected to be larger than for unrelated sequences, where $n_0$ corresponds to coincidental mutations.
(B)~Positive and negative pairs are called mutated if both sequences have mutations $n_1,n_2>0$. Among positive pairs in the synthetic datasets, more than $99\%$ are mutated. 
(C, D) Distributions of the rescaled variables $x'$ and $y$ (\ref{xyeq}), for pairs of synthetic sequences belonging to the same lineage (positive pairs) and sequences belonging to different lineages (negative pairs). The separatrix $x'-y=t'$ marks a high-precision (99\%) threshold choice.
(E)~To limit the number of pairwise comparisons we make use of high-precision and high-sensitivity CDR3-based partitions. High precision corresponds to the choice $t=t^*$. High sensitivity corresponds to a coarser partition where $t$ is set to achieve 90\% sensitivity. When the two partitions disagree, mutational information can be used to break the coarse, high-sensitivity partition into smaller clonal families.
(F,G)~Mutations-based methods achieve high sensitivity across all CDR3 lengths $l$ in the synthetic dataset~(G), extending the range of applicability with respect to the CDR3-based method~(F).}

\newcommand{\figthreeSonetitle}{\TM Merging partitions.}
\newcommand{\figthreeSonecaption}{\TM Red circles represent clusters from the coarse
            (high-sensitivity) partition, while green clusters
            represent the fine (high-precision) partition. When the
            two partitions differ, HILARy-full merges precise clusters
            inside each sensitive cluster whenever there exists of pair
            of positive sequences linking them.}
\newcommand{\figthreeStwotitle}{\TM Error vs \VJl class size.}
\newcommand{\figthreeStwocaption}{\TM We plot the fitting
            error of $P(x)$
            by the mixture model, for each \VJl class in the synthetic
            dataset, as a function of
            their sizes. The error is computed as the squared
            difference between the model and data distributions of distances.}

\ifthenelse{\boolean{elife}}{
\begin{figure}
\includegraphics[width=\linewidth]{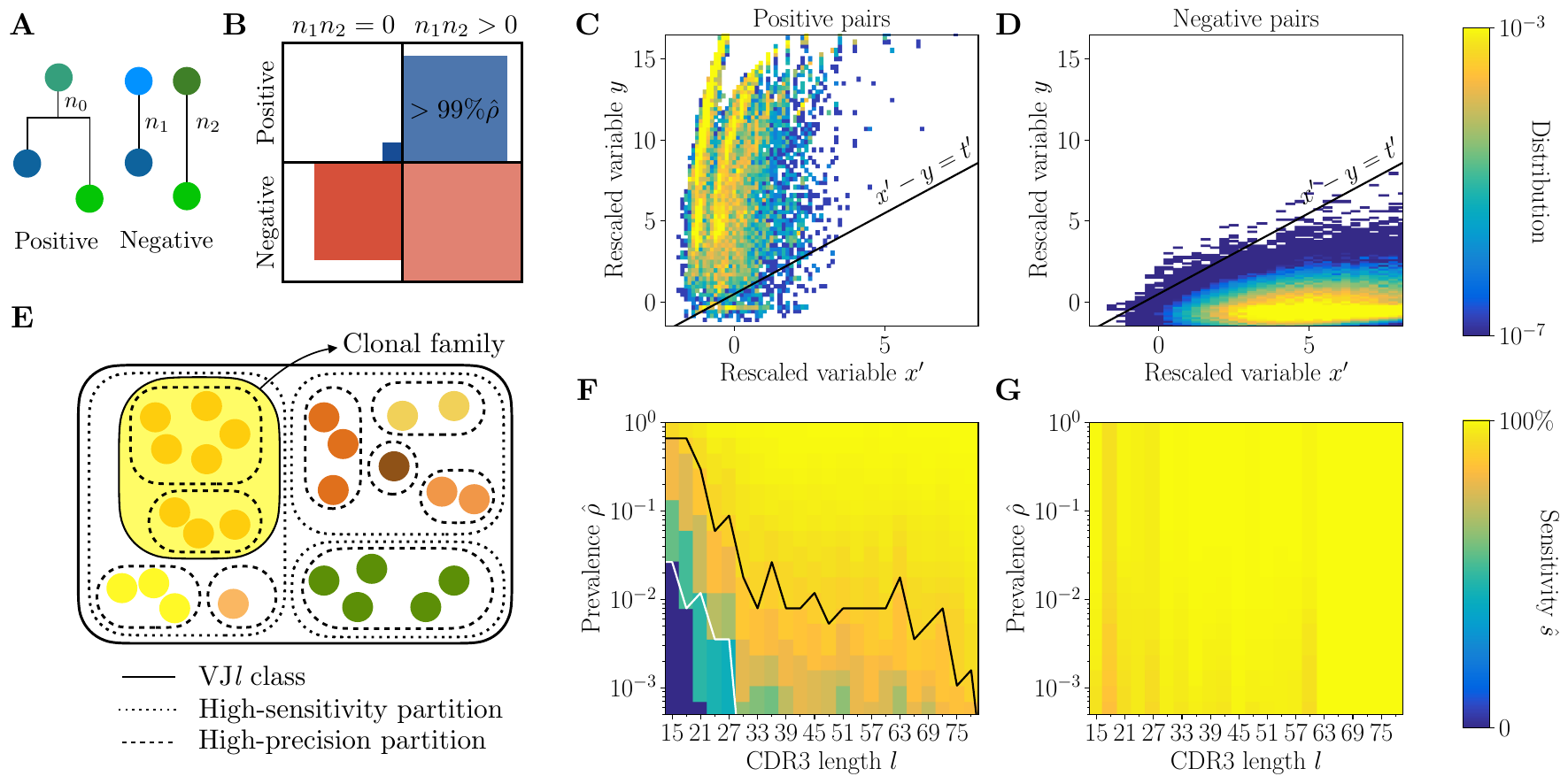}
\caption{\figthreecaption}
\label{fig:figure3}
\figsupp[{\figthreeSonetitle}]
{{\bf \figthreeSonetitle} \figthreeSonecaption}
{\includegraphics[width=\linewidth]{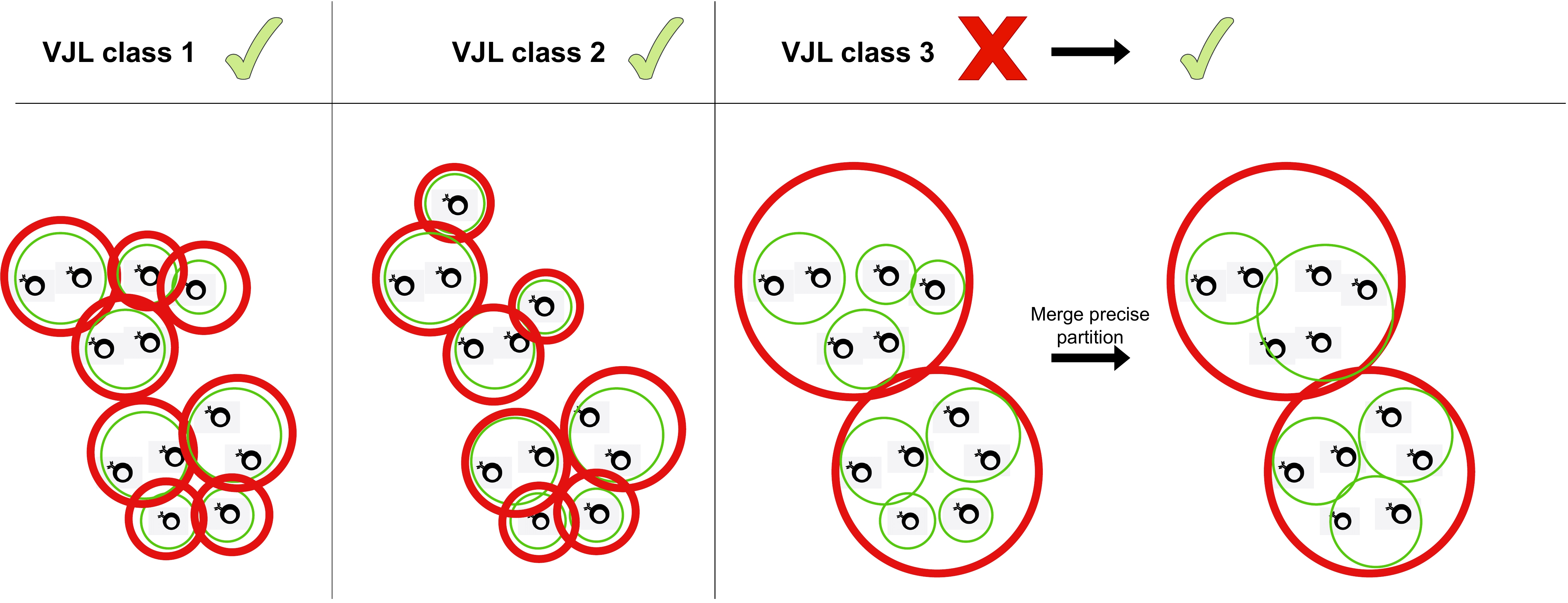}}
\label{figsupp:figure8}
\figsupp[{\figthreeStwotitle}]
{{\bf \figthreeStwotitle} \figthreeStwocaption}
{\includegraphics[width=\linewidth]{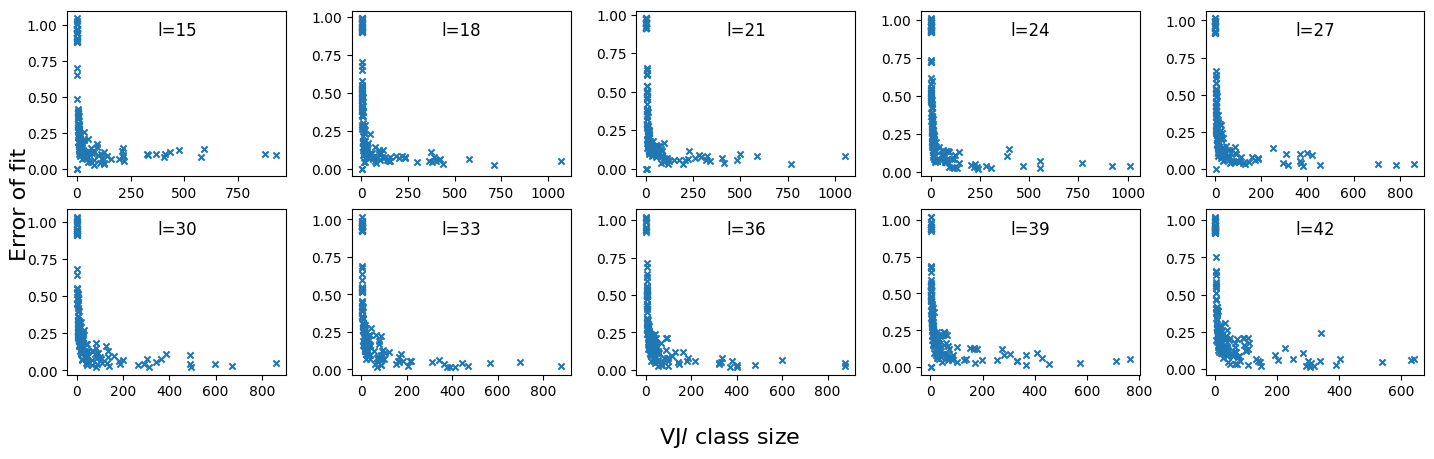}}
\label{figsupp:figure9}
\end{figure}
}
{
\begin{figure*}
\includegraphics[width=\linewidth]{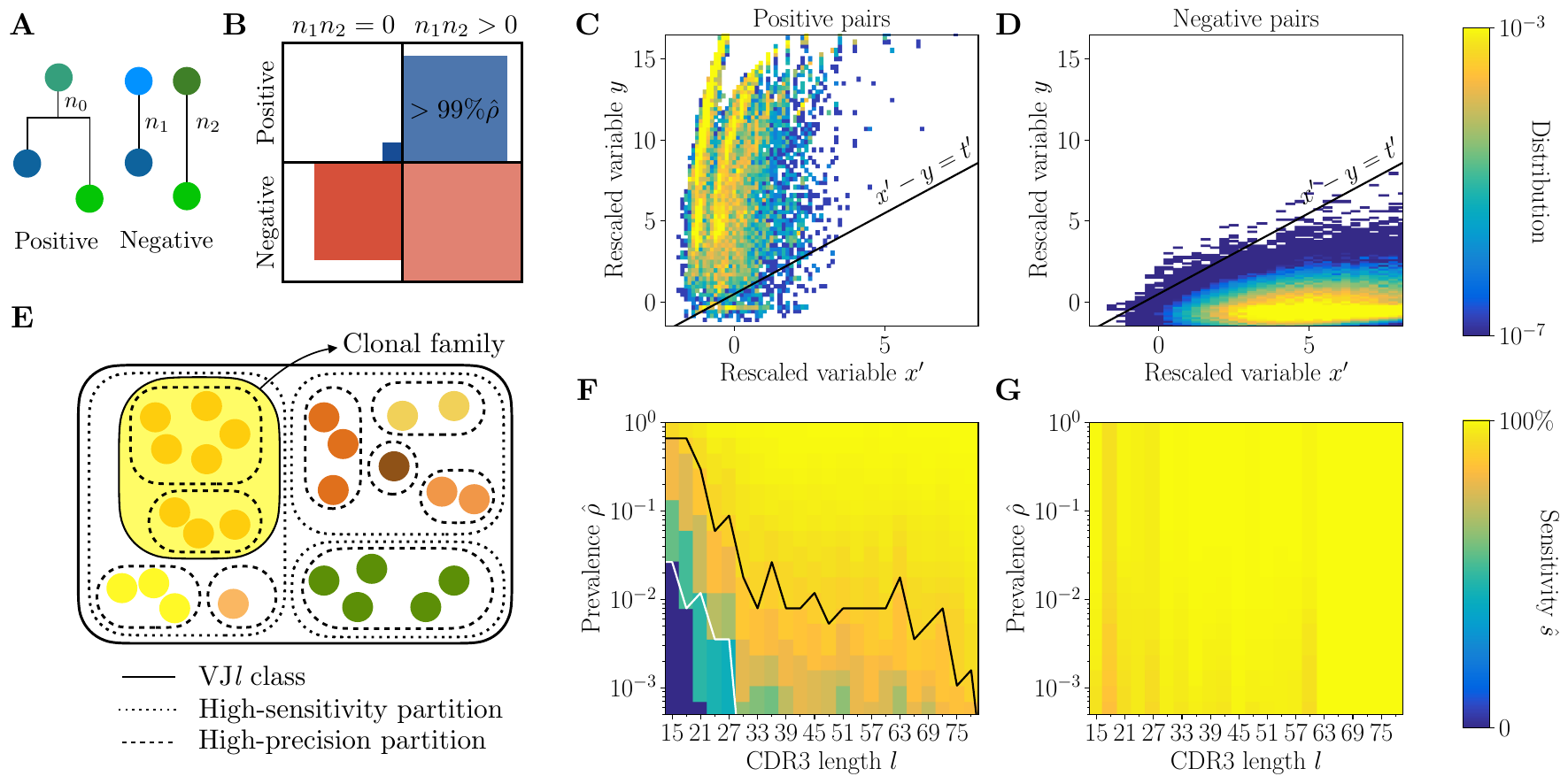}
\caption{\figthreecaption}
\label{fig:figure3}
\end{figure*}
}

\subsection{Tests on synthetic datasets} 
So far we have presented a method to set a high-precision threshold with predictable sensitivity, based on estimates from the distribution of distances $P(x)$ only. 
To verify that these performance predictions hold in {\GA a realistic} inference task, we designed a method to generate realistic synthetic datasets where the clonal family structure is known. This generative method will also be used in the next sections to create a benchmark for comparing different clustering algorithms.

We first estimated the distribution of clonal family sizes from the data of \cite{Briney2019} by applying {\TM HILARy-CDR3} with adaptive threshold as described above to \VJl classes for which the inference was highly reliable, i.e. for which the predicted sensitivity was $\hat s(t^*)\geq 90\%$. In that limit, clusters are clearly separated and the partition should depend only weakly on the choice of clustering method. The resulting distribution of clone sizes follows a power-law with exponent $-2.3$.

{\TM To create a synthetic lineage, we first
  draw a random progenitor using the soNNia model for IgH generation (\SIFIG{figure2}{figure4}). We then draw the size of the lineage at random, using the power-law distribution above. Mutations are then randomly drawn on each sequence of the lineage} in a way that preserves the mutation sharing patterns observed in families of comparable size from the partitioned data (\SIFIG{figure2}{figure5}). We thus generated $10^4$ lineages and $2.5\cdot 10^4$ sequences. {\TM Note that, while that procedure is partially based on real data, in particular the distribution of lineage sizes and mutational co-occurence structure in the lineages, it uses completely random sequences and mutations. In addition, these empirical observables were inferred from \VJl classes that were easy to cluster, ensuring that they are not biased by our inference method, and therefore should not give it an unfair advantage.}
More details about the procedure are given in the \METH{Methods-synthetic}.

We applied the {\TM HILARy-CDR3} method to this synthetic dataset. The sensitivity achieved at $t^*$ roughly follows and sometimes even outperforms the predicted one $\hat s(t^*)$ across different values of $\rho$ and $l$ (\FIG{figure2}J, dashed line), validating the approach and the choice of the adaptive high-precision threshold $t^*$ (the discrepancy is due to the fact that $\mu$ is assumed to be constant in the prediction, while it varies in the dataset). These results also confirm the poor performance of the method at low prevalences and short CDR3s.

\subsection{Incorporating phylogenetic signal} 
\label{Results-phylogenetic}
To improve the performance of {\TM HILARy-CDR3}, we set out to include the phylogenetic signal encoded in the mutation spectrum of the templated regions of the sequences.
Two sequences belonging to the same lineage are expected to share some part of the mutational histories, and therefore sequences with shared mutations are more likely to be in the same lineage.

We focus on the template-aligned region of the sequence outside of the CDR3, where we can reliably identify substitutions with respect to the germline. We denote the length of this alignment by $L$, so that the total length of the sequence is $l+L$. 
For each pair of sequences, we define $n_1,n_2$ as the number of mutations along the templated alignment in the two sequences, $n_0$ the number of mutations shared by the two, and $n_L=n_1+n_2-2n_0$ the number of non-shared mutations. Under the hypothesis of shared ancestry, the $n_0$ shared mutations fall on the shared part of the phylogeny, and are expected to be more numerous than under the null hypothesis of independent sequences, where they are a result of random co-occurrence (\FIG{figure3}A). 

To balance the tradeoff between the information encoded in the templated part of the sequence and the recombination junction, we can compute characteristic scales for the two variables of interest:
the number of shared mutations $n_0$ and the CDR3 {\TM distance} $n$. 
Intuitively, in highly mutated sequences, we can expect substantial divergence in the CDR3. At the same time, the number of mutations in the templated regions would increase, possibly leading to more shared mutations. Conversely, sequences with few or no mutations carry no information in the templated region, but we also expect their CDR3 sequences to be nearly identical. To adapt a clustering threshold to the two variables, we compute their expectations under the two assumptions, and define the rescaled variables
{\TM
\begin{equation} 
x'=\frac{n-\langle n\rangle_{\rm T}}{\sigma_{\rm T}(n)},\quad y=\frac{n_0-\langle n_0\rangle_{\rm F}}{\sigma_{\rm F}(n_0)},
\label{xyeq}
\end{equation} 
where $\langle n \rangle_{\rm T}=l(n_L+1)/L$ is the expected value of $n$} under the hypothesis that sequences belong to the same lineage {\TM(see Methods)}, and {\TM$\langle n_0\rangle_{\rm F}=n_1n_2/L$} is the expected value of $n_0$ under the hypothesis that they do not. The standard deviations are likewise defined as {\TM$\sigma_{\rm T}(n) = \sqrt{ \langle n^2\rangle_{\rm T}- \langle n \rangle_{\rm T}^2}=(1/L)\sqrt{l(l+L)(n_L+1)}$} and {\TM$\sigma_{\rm F}(n_0) = \sqrt{ \langle n_0^2\rangle_{\rm F}- \langle n_0 \rangle_{\rm F}^2}=\sqrt{n_1n_2/L}$} (\METH{Methods-mutations}).

For more than $99\%$ of positive pairs, both sequences are mutated, i.e. $n_1,n_2>0$ (\FIG{figure3}B). Without loss of sensitivity, we focus on the mutated part of the dataset, since we cannot use $y$ for non-mutated sequences. The distributions of $x'$ and $y$ for positive and negative pairs (\SIFIG{figure2}{figure3}C and~D) are well separated, with positive pairs characterized by an overrepresentation of shared mutations. By adding the phylogenetic signal $y$ we can identify positive pairs of sequences that have significantly diverged in their CDR3 ($x'>0$) but share significantly more mutations than expected (large $y$).

Computing $y$ for each pair of sequences is {\TM computationally expensive. To avoid examining all pairs, we first perform two different nested clusterings of each \VJl class using the CDR3-based method: the previously described HILARy-CDR3 ``fine'' partition with threshold $t^*$ that ensures high precision $\hat \pi=99\%$; and a ``coarser'' clustering with a high threshold $t=t_{\rm sens}$ that ensures high estimated sensitivity $\hat s=90\%$ (\METH{Methods-CDR3} and \FIG{figure3}E).
When lineages are easily separable (e.g. for sufficiently large prevalence $\rho$ and CDR3 length $l$), these two partitions coincide, and we do not need to compute $y$ at all. When they do not coincide, we can use the phylogenetic signal $y$ to refine the coarse high-sensitivity partition. We only need to compute $y$ for pairs that belong to the same coarse cluster, but not to the same fine cluster: the phylogenetic signal $y$ is used to merging the fine-partition clusters into clonal families (\METH{Methods-mutations} and \SIFIG{figure3}{figure8}).} This allows us to considerably reduce the number of pairwise comparisons that we need to make between the templated regions of the sequences.

Using $x'$ and $y$, we classify pairs of sequences as positive (i.e. belonging to the same family) if { \GA $y\ge x'-t'$, and as negative otherwise}.
We {\TM can} compute the expected sensitivity on the synthetic data, and find that it reaches values $\geq 90\%$ across the whole range of prevalence $\rho$ and CDR3 lengths $l$, outperforming {\TM HILARy-CDR3} in the low-$\rho$, low-$l$ region (\FIG{figure3}F and G). This proves that using the phylogenetic signal significantly improves performance over {\TM HILARy-CDR3.}

{\TM The procedure outlined above, which we call HILARy-full, may be summarized as follows: (1) group sequences by \VJl class; (2) apply HILARy-CDR3 twice, once with the high-precision threshold as before to get a fine partition, and once with a high-sensitivity threshold to get a coarse partition, thus obtaining two nested partitions; (3) compute $x'$ and $y$ using Eq.~\ref{xyeq} only for pairs that belong to the same coarse cluster but to different fine clusters; (4) merge all fine clusters with at least one pair with {\GA $y\geq x'-t'$.}}

\newcommand{\figfourcaption}{\TM
{\bf Benchmark of the alternative methods on synthetic heavy-chain repertoires.}
(A) Comparison of inference time using subsamples from the largest \VJl class found in donor 326651 from \cite{Briney2019}. Comparisons were done on a computer with 14 double-threaded 2.60GHz CPUs (28 threads in total) and 62.7Gb of RAM. (B) Clustering precision $\pi_{\text{post}}$ (post single linkage clustering of positive pairs), (C) sensitivity $s_{\text{post}}$, and (D) variation of information $v$ across CDR3 length $l$ in the realistic synthetic dataset generated for this study.
 Solid lines represent the mean value averaged over 5 synthetic datasets.
 (E-G): Same as (B-D) but for the synthetic dataset from \cite{ralph2022inference} designed for the development and testing of the partis software. The solid lines represent the mean over the three {\GA datasets}.}

\newcommand{\figfourSonetitle}{\TM Method comparison as a function of mutation rate.}
\newcommand{\figfourSonecaption}{\TM Synthetic data from the benchmark are taken from \cite{ralph2022inference}.
    (A) Clustering precision $\pi_{\text{post}}$ (post single linkage clustering of positive pairs), (B) sensitivity $s_{\text{post}}$, and (C) variation of information $v$ vs mutation rates using the heavy chain only.
    Solid lines represent the mean value averaged over the 3 datasets. (D-F): Same than (A-C) using paired light and heavy chains.}
  
\newcommand{\figfourStwotitle}{\TM Performance of single-linkage
            clustering with fixed threshold.}
\newcommand{\figfourStwocaption}{\TM We call this method VJCDR3-sim,
            where sim is the threshold on the normalized similarity between two CDR3s,
            equal to $1-x$, where $x$ is our normalized Hamming distance.
       (A) Clustering precision $\pi_{\text{post}}$ (post single linkage clustering of positive pairs), (B) sensitivity $s_{\text{post}}$, and (C) variation of information $v$ across CDR3 length $l$ in the realistic synthetic dataset generated for this study.
          Solid lines represent the mean value averaged over 5 synthetic datasets. (D-F): Same than (A-C) using the synthetic data from \cite{ralph2022inference} and across mutation rates.}
        
\newcommand{\figfourSthreetitle}{\TM Performance of spectral SCOPer using V gene mutations.}
\newcommand{\figfourSthreecaption}{\TM (A) Comparison of inference time using subsamples from the largest \VJl class found in donor 326651 from \cite{Briney2019}. Comparisons were done on a computer with 14 double-threaded 2.60GHz CPUs (28 threads in total) and 62.7Gb of RAM. (B) Clustering precision $\pi_{\text{post}}$ (post single linkage clustering of positive pairs), (C) sensitivity $s_{\text{post}}$, and (D) variation of information $v$ vs CDR3 length $l$ in the realistic synthetic dataset generated for this study. Solid lines represent the mean value averaged over 5 synthetic datasets.}

\ifthenelse{\boolean{elife}}{
\begin{figure}
\includegraphics[width=\linewidth]{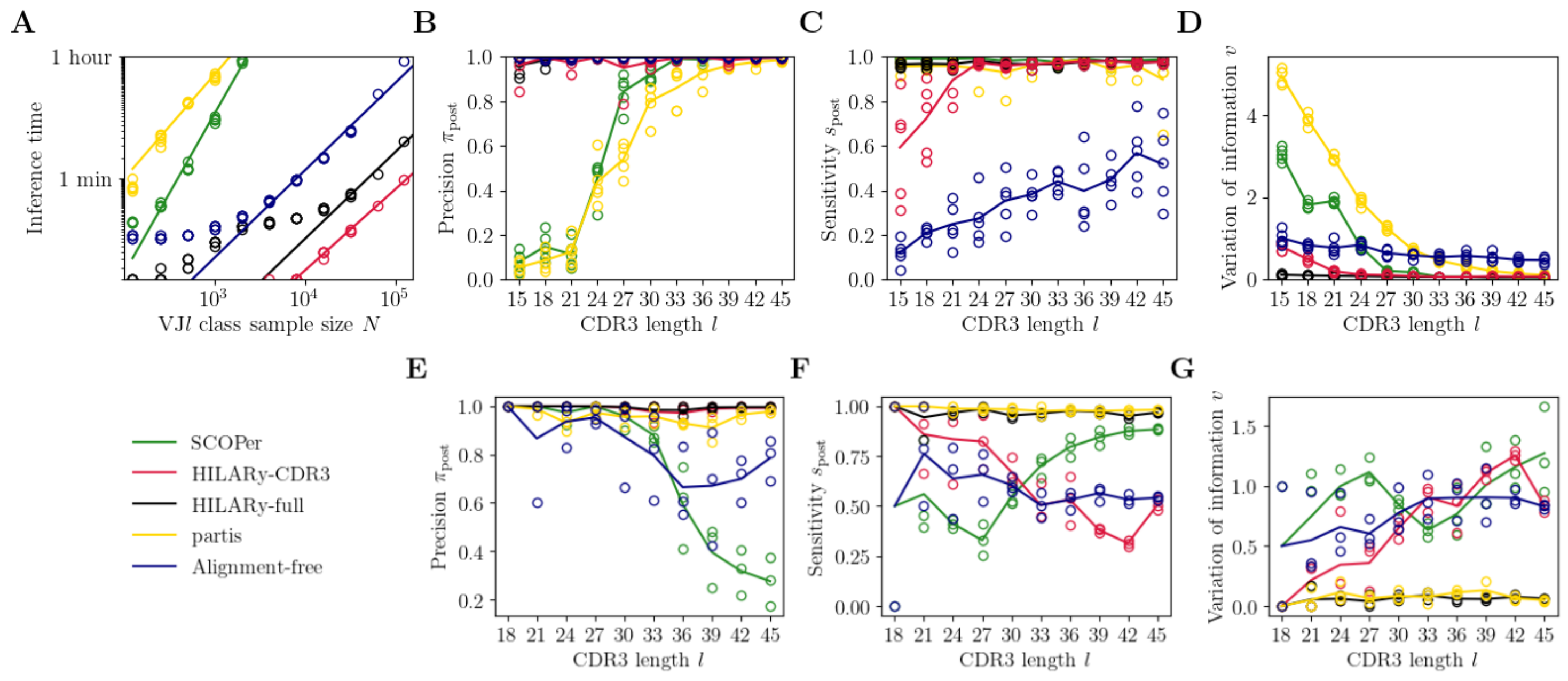}
\caption{\figfourcaption}
\label{fig:figure4}
\figsupp[{\figfourSonetitle}]
{{\bf \figfourSonetitle} \figfourSonecaption}
{\includegraphics[width=\linewidth]{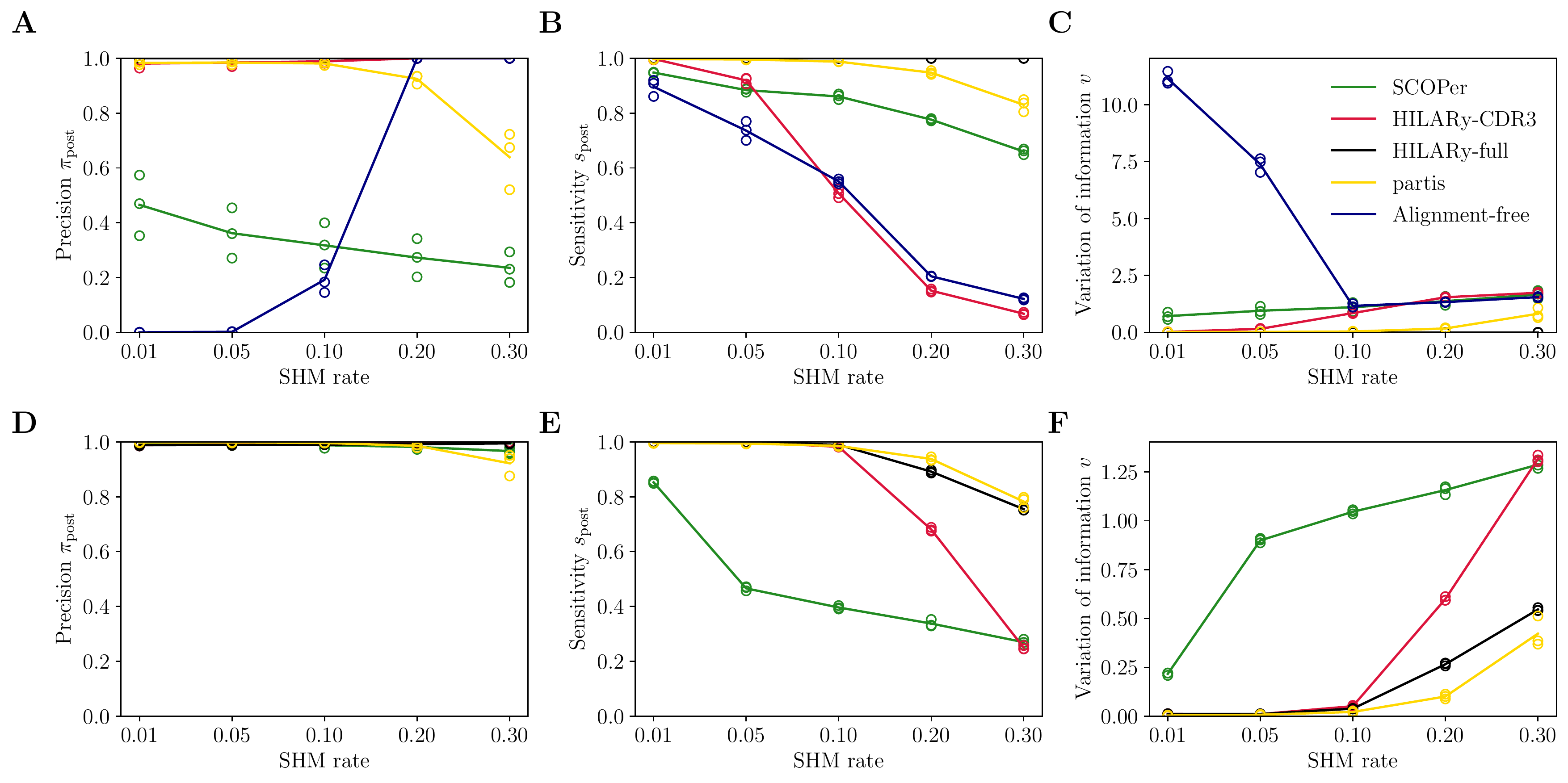}}
\label{figsupp:figure11}
\figsupp[{\figfourStwotitle}]
{{\bf \figfourStwotitle} \figfourStwocaption}
{\includegraphics[width=\linewidth]{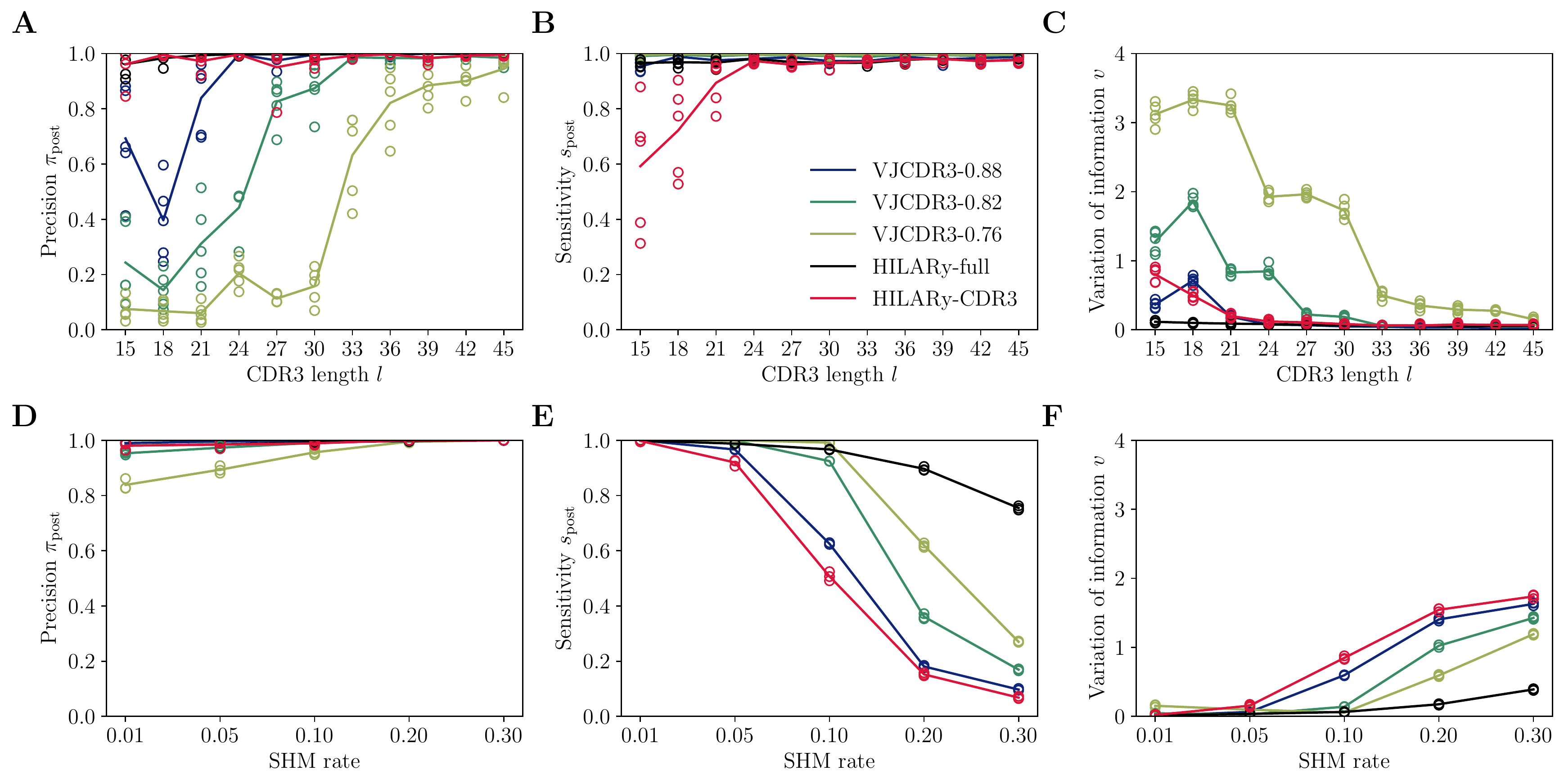}}
\label{figsupp:figure10}
\figsupp[{\figfourSthreetitle}]
{{\bf \figfourSthreetitle} \figfourSthreecaption}
{\includegraphics[width=\linewidth]{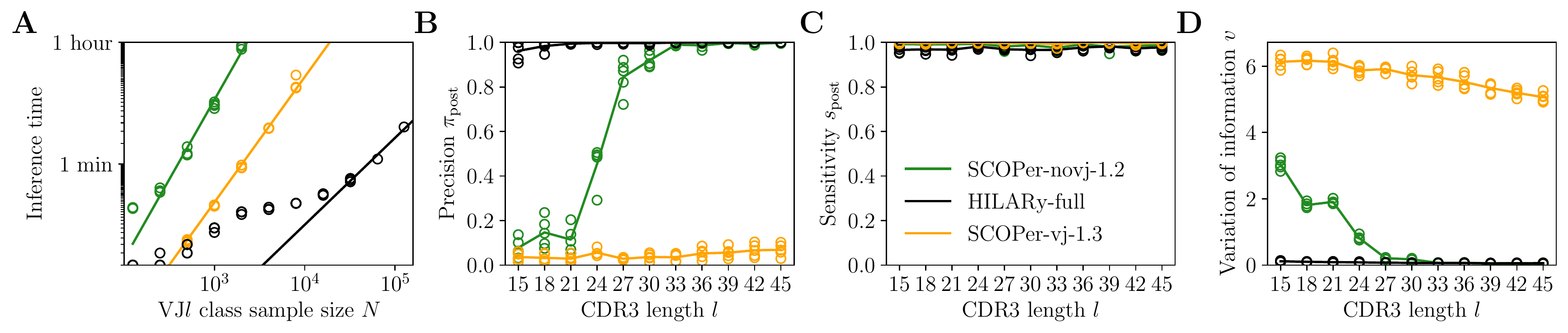}}
\label{figsupp:figure12}
\end{figure}
}
{
\begin{figure*}
\includegraphics[width=\linewidth]{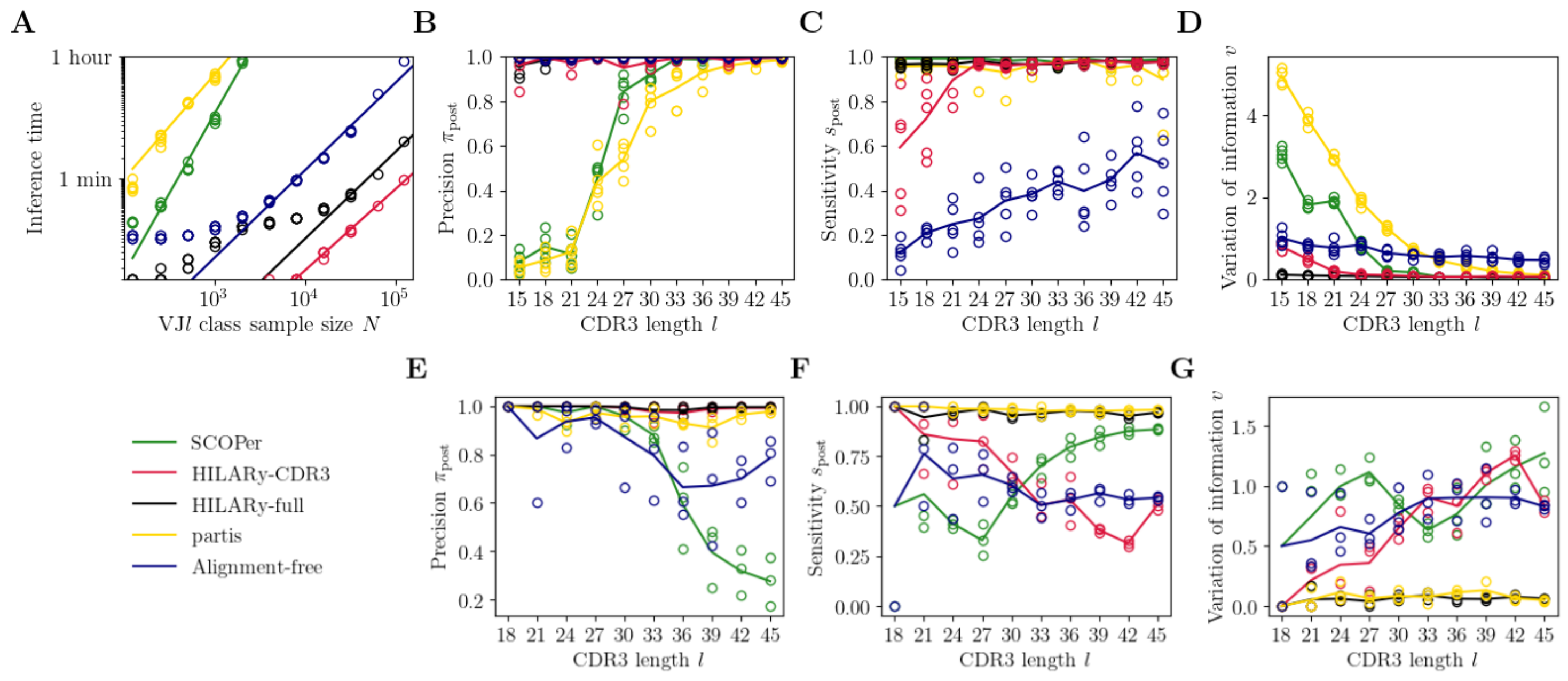}
\caption{\figthreecaption}
\label{fig:figure4}
\end{figure*}
}

\begin{figure*}
	\includegraphics[width=\linewidth]{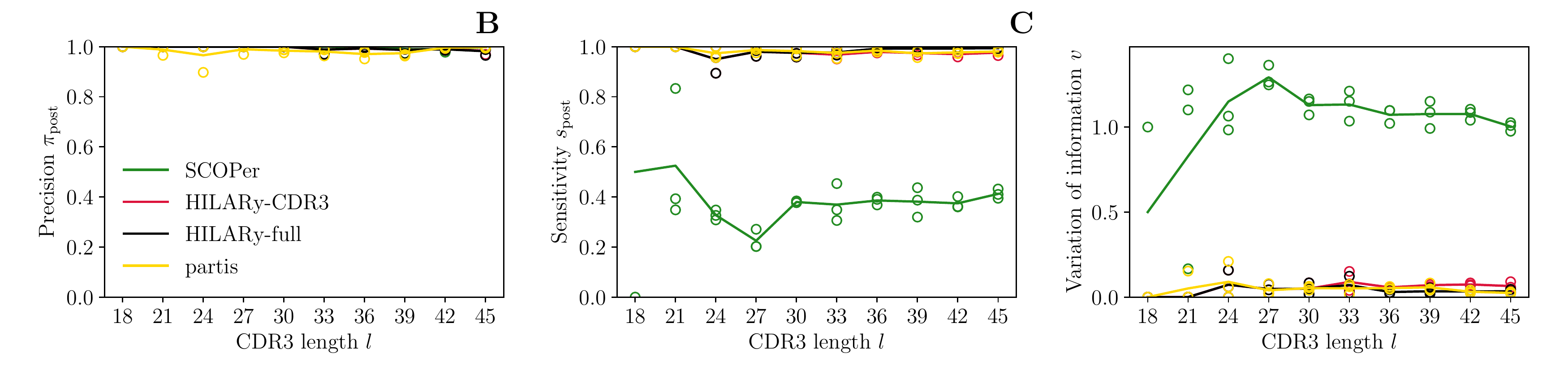}
	\caption{\TM
		{\bf Benchmark of HILARy with paired light and heavy chains.} 
		 (A) Clustering precision $\pi_{\text{post}}$ (post single linkage clustering of positive pairs), (B) sensitivity $s_{\text{post}}$, and (C) variation of information $v$ across CDR3 length $l$, on the synthetic datasets from \cite{ralph2022inference} designed for the development and testing of the partis software.
	}
\label{fig:figure5}  
\end{figure*}

\subsection{Benchmark of the methods} 

We compare our approach to state-of-the-art methods. {\TM In addition to our two algorithms---HILARy-CDR3 and HILARy-full---our benchmark includes the alignment-free method of \cite{lindenbaum2021alignment}, partis \citep{ralph2016LikelihoodBasedInferenceCell}, {\TM and the spectral clustering method of  SCOPer \citep{nouri2018spectral}. {\GA The SCOPer method using V and J gene mutations \cite{nouri2020somatic} was also tested, but gave worse results (\SIFIG{figure4}{figure12}).} Details about the used versions and parameters are referenced in the data availability section.} We tested all algorithms on two synthetic datasets: a dataset simulated by the partis package and used in \cite{ralph2022inference} to benchmark partis against increasing levels of somatic hypermutations; and the synthetic data described above.
That dataset is more realistic in the sense that it represents well the statistics of mutation patterns and, perhaps more importantly, the long-tail distribution of clone sizes observed in the data, with its large impact on the diversity of prevalences, which play an important role in the inference. The partis dataset is generated from a population genetics model. It provides a more independent test since it is not based on data used to develop the method and allows to study performance across different mutation rates.}

First, we measure the inference time of each algorithm {\TM on our synthetic data set}. We find that the inference time is primarily affected by the size of the largest \VJl class. Therefore, we measure the inference time using the largest class found in donor 326651 of \cite{Briney2019} with the size of $N=1.2\times 10^{5}$ unique sequences. We then apply the methods to a series of subsamples of this class to get the computational time as a function of the subsample size (\FIG{figure4}A). We only allowed for runtimes below 1~hour. We find that only 3 methods achieve satisfactory performance {\TM(under an hour)}: the two methods introduced here, and the alignment-free method. The other two methods, SCOPer and partis, are limited to \VJl classes of small size {\TM ($<10^4$ and $<10^3$ respectively)}.

To compare the five algorithms in finite time, we test the accuracy of the methods using synthetic datasets with different CDR3 lengths, {\TM and with fixed mutation rate of {\GA10\%} for the partis dataset (the mutation rate is not adjustable in our synthetic dataset as it mimics that of the data)}. We focus on short CDR3s, $l\in[15,45]$, which are the most challenging for lineage inference. Each dataset contains $10^4$ unique sequences, so that the dominant \VJl class is typically of size $\sim10^3$ and can be handled by all 5 algorithms. 
We measure performance using three metrics applied to the resulting partition: pairwise sensitivity $s_{\text{post}}$ (\FIG{figure4}B,E), pairwise precision $\pi_{\text{post}}$ (\FIG{figure4}C,F), and the variation of information $v$ (\FIG{figure4}D,G). {\TM Performance measures as a function of mutation rate in the partis dataset are presented in \SIFIG{figure4}{figure11}A-C.}
Pairwise sensitivity $s_{\text{post}}$ and precision $\pi_{\text{post}}$ are \emph{a posteriori} analogs of the a priori estimates defined before in Eqs.~\ref{precision} and \ref{sensitivity}, now computed after propagating links through the transitivity rule of single linkage clustering. Their value reflects not only the accuracy of the adaptive threshold but is also affected by the propagation of errors in single linkage clustering.
Variation of information is a global metric of clustering performance which measures the loss of information from the true partition to the inferred one, and is equal to zero for perfect inference and positive otherwise (\METH{Methods-evaluation}).

\begin{figure*}
\includegraphics[width=\linewidth]{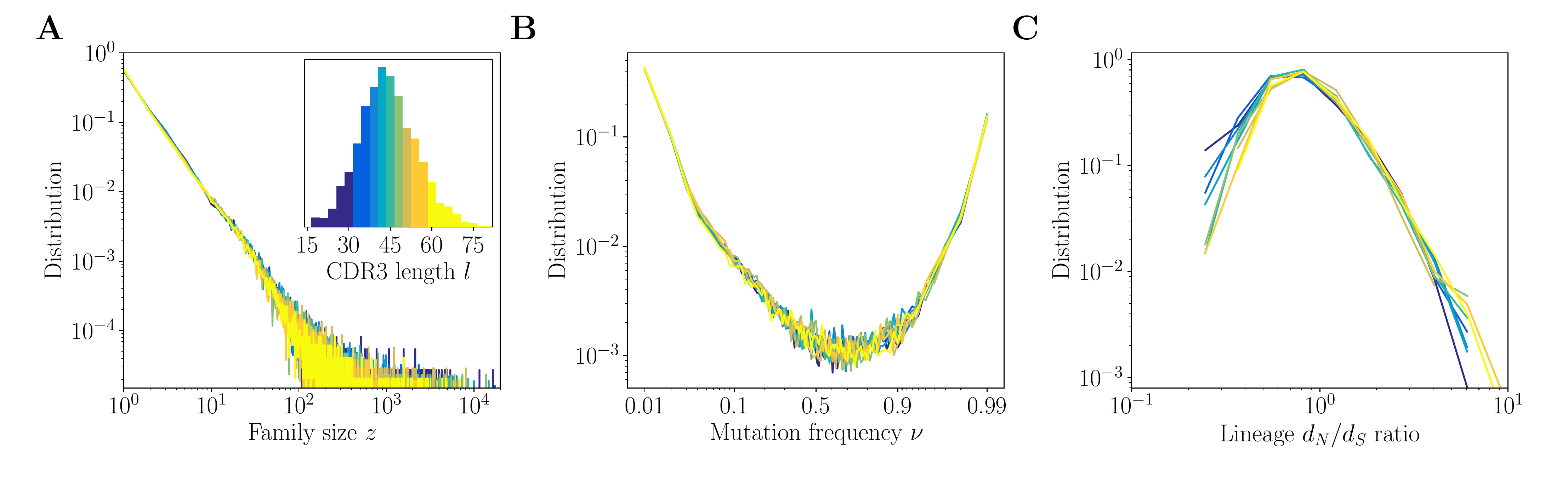}
\caption{
{\bf Inference results across CDR3 lengths.}
Inference results for donor 326651 of \cite{Briney2019} are presented for 9-quantiles of the CDR3 distribution, each containing between $8$ and $12\%$ of the total number of sequences (corresponding to 9 colors in the inset of A). 
(A) Distributions of family size $z$. All CDR3 length quantiles exhibit universal power-law scaling with exponent~$-2.3$ 
(B) Site frequency spectra estimated for families of sizes $z=100$. Families of larger sizes were subsampled to $z=100$ to subtract the influence of varying family sizes.
(C) Distribution of lineage $d_N/d_S$ ratios computed for polymorphisms in CDR3 regions over all lineages within each 9-quantile. 
}
\label{fig:figure6}  
\end{figure*}

We find that, out of the five tested methods, only our method achieved both high sensitivity and high precision across all CDR3 lengths {\TM and for both datasets}. The {\TM HILARy-CDR3} method achieves high precision everywhere by construction, but only reached good sensitivity for CDR3 lengths 24 and above. The alignment-free method also achieves high precision everywhere, but with low sensitivity, meaning that it erroneously breaks up clonal families into smaller subsets. These three methods achieve good precision thanks to the use of a null model for the negative pairs.
On the contrary, SCOPer has excellent sensitivity everywhere but only achieves high precision for large lengths ($l>30$), suggesting that it erroneously merges short-CDR3 clonal families. Likewise, partis has {\TM high} sensitivity but loses precision {\TM for short CDR3 on our realistic dataset}, meaning that many clonal families are erroneously merged again. {(\TM Note that our definition of precision and sensitivity differs from those used in \cite{ralph2022inference}, which explains the differences between the performance measures reported here and in \cite{ralph2022inference}.)}

{\TM The variation of information offers a useful summary of the performance (\FIG{figure4}D and G). According to that measure, only HILARy-full performs well across parameters and datasets.}
We conclude that {\TM HILARy-CDR3} should be chosen for its consistently high sensitivity, specificity, and speed. In the case of the largest datasets, the faster {\TM HILARy-CDR3} is a useful alternative for long enough CDR3s {\TM in realistic repertoires}.

\subsection{\TM Extension to heavy/light chain paired data} 
{\TM We added an extension of HILARy to infer lineages from paired chain repertoires, i.e. with paired light and heavy chain sequences. To extend HILARy-CDR3, we generalize the \VJl class to a V$_H$ J$_H$ V$_L$ J$_Ll_{H+L}$ class, using the V and J genes from both the heavy and light chains, and the sum of their CDR3 lengths $l_{H+L}=l_H+l_L$. We then apply HILARy-CDR3 using the sum of the Hamming distances between the heavy and light chain CDR3s, normalized by $l_H+I_L$, as our new paired-chain $x$. The null distribution used is computed with soNNia using a default generation model for paired heavy and light chains. We incorporate the phylogenetic signal of both chains by concatenating  their respective template genes, to obtain the total mutation counts $n_0 = n_{0,H}+n_{0,L}$, and using $L$ as the sum of the lengths of the V$_{H}$ and V$_L$ genes.

In \FIG{figure5} we compare our method to SCOPer and partis on the synthetic dataset from \cite{ralph2022inference} as a function of CDR3 length, as our method for generating synthetic sequences could not be easily extended to add random light chains. Performance comparison as a function of mutation rate is presented in \SIFIG{figure4}{figure11}D-F. HILARy performs better than SCOPer and comparably to partis, which was designed and tested against this dataset.}

\subsection{Inference of clonal families in a healthy repertoire} 
We next use our method to infer the clonal families of the heavy-chain IgG repertoires of healthy donors from \cite{Briney2019}. \FIG{figure6} summarizes key properties of the inferred clonal families of donor 326651. 
We take advantage of the consistency of our method across CDR3 lengths, as evidenced by the benchmark, to study how the lineage structure changes with the CDR3 length. To this end, we divide the dataset into 9 quantiles, each containing {\GA $\sim 10\%$} of the total number of sequences (\FIG{figure6}A, inset).

We find that across the 9 subsets of the data, the statistics of the lineage structure inferred with the mutations-based method are largely universal. The distribution of the clonal family sizes $z$ (\FIG{figure6}A) follows a power law across all CDR3 lengths under study, with no significant differences between different lengths. This results generalizes an earlier observation used above for generating synthetic datasets, but which was restricted to high-$\hat \rho$, high-$l$ \VJl classes, and justifies {\em a posteriori} the use of a universal power law in the generative model.

For the largest families, of size $z\ge100$, we compute two intra-lineage summary statistics: the site frequency spectrum, which gives the distribution of frequencies of point mutations within lineages, and the distribution of $d_N/d_S$ ratios between non-synonymous and synonymous CDR3 polymorphisms within clonal families {\TM(estimated by counting)}. To avoid the bias of the varying family sizes, we subsampled all families to size $z=100$.

Under models of neutral evolution with fixed population size, the distribution of point-mutation frequencies $\nu$ goes as $\nu^{-1}$. Here we observe a non-neutral profile of the spectrum, with an upturn at large allele frequencies $\nu>0.5$ (\FIG{figure6}B). It is a known signature of selection or of rapid clonal expansion \cite{Horns2019, Nourmohammad2019}. We find that site frequency spectra are universal for all CDR3 lengths, suggesting that the dynamics that give rise to the structure of lineages and the subsequent dynamics that influence the sampling of family members, do not depend on the CDR3 length.

The lineage $d_N/d_S$ ratio is also largely consistent across CDR3 lengths (\FIG{figure6}C), while spanning 2 orders of magnitude, suggesting a wide gamut of selection forces. We could have expected longer loops to be under stronger purifying selection (lower $d_N/d_S$) to maintain their specificity and folding.
Instead, we observe that short CDR3s have more lineages with low $d_N/d_S$. This may be due to different sequence context and codon composition in short versus long CDR3s. Short junctions are largely templated, whereas long junctions have long, non-templated insertions, and it was shown that templated regions have evolved their codons to minimize the possibility of non-synonymous mutations \citep{Saini2015}, which would lead to a lower $d_N/d_S$, regardless of selection.

\section{Discussion}

Clonal families are the building blocks of memory repertoire shaped by VDJ recombination and subsequent somatic hypermutations and selection. Repertoire sequencing datasets enable new approaches to understand these processes. They allow us to model the different sources of diversity and measure the selection pressures involved. To take full advantage of this opportunity we need to reliably identify independent lineages.

Here, we introduced a general framework for studying the methods for partitioning high-throughput sequencing of BCR repertoire datasets into clonal families. We have identified the main factors that influence the difficulty of this inference task: low clonality levels and short recombination junctions. We quantified the clonality level using the definition of pairwise prevalence $\rho$ and introduced a method to estimate it a priori, without knowing the partition. We found the prevalence levels across \VJl classes to span three orders of magnitude (\FIG{figure2}B), unraveling the varying degree of complexity.

We leveraged the soNNia model of VDJ recombination to quantify the CDR3 diversity and constructed a null expectation for the divergence of independent recombination products. This null model enabled the design of a CDR3-based clustering method with an adaptive threshold, {\TM HILARy-CDR3}, that allows us to keep the precision of inference high across prevalences and CDR3 lengths. Owing to the prefix tree representation of the CDR3 sequences, this method is characterized by very short inference times thanks to avoiding all pairwise comparisons in single linkage clustering. As expected, we found that the adaptive threshold choice limits the sensitivity of inference in the regime of short junctions and low prevalence (\FIG{figure3}F, {\TM below the black line}). 

To remedy the limitations of the CDR3-based approach we developed a mutation-based method {\TM (HILARy-full)}. We found that including the phylogenetic signal of shared mutations in highly mutated sequences allows us to properly classify them into lineages despite significant CDR3 divergence. We studied the performance of the method using synthetic data and found significant improvement with respect to {\TM HILARy-CDR3}: we extended the range of high-precision and high-sensitivity performance to cover all values of prevalence and CDR3 lengths observed in productive data (\FIG{figure3}G). 

We have compared the two methods developed here with state-of-the-art approaches: the partis \citep{ralph2016LikelihoodBasedInferenceCell} and {\GA SCOPer \citep{nouri2018spectral} }algorithms, and the alignment-free method \citep{lindenbaum2021alignment}. {\TM Compared to these methods, HILARy relies on a probabilistic model of VDJ recombination and selection, which allows it to explicitly control for precision. This is not possible in partis, which relies on likelihood ratio test to merge candidate clusters together to form families. SCOPer also chooses a clustering threshold based on the pairwise distribution of distances, but without a null model.
Another innovation of HILARy-full is to use a null expectation for the number of shared mutations. This feature makes the method robust to varying levels of mutation rates across sequences.
HILARy achieves optimal efficiency by combining CDR3-based and mutation-based information. Typically, a large part of the dataset doesn’t require the use of the full method, allowing for greatly reduced inference times.
HILARy relies on the soNNia model, which is based on a neural network}, and benefits from its expressivity to quantify the purifying selection that modifies the VDJ recombination statistics. We found the performance of this model satisfactory when applied to healthy memory repertoires, in agreement with previous findings \citep{Isacchini2021, ortega2021modeling}. However, purifying selection is expected to be more pronounced in datasets of disease-specific cohorts and a default soNNia model may overestimate the diversity \citep{Mayer2022} and lead to underestimation of the fallout rate. The inference framework introduced here could still be applied with more sophisticated models of selection, and take advantage of higher levels of clonality that characterize many disease-specific datasets \citep{nielsen2020human, turner2020human}.

We applied the mutations-based method to infer lineages in a repertoire of a healthy donor, sequenced at great depth \citep{Briney2019}. We took advantage of the consistency our method exhibits across CDR3 lengths to find that the statistics of lineages, including a heavy-tail distribution of family sizes as well as signatures of selection, are universal and independent of the CDR3 length. This result implies that the dynamics of expansion, mutation, and selection are independent of the CDR3 and suggests they are dictated by the rules of affinity maturation and memory formation rather than BCR specificity. It advocates for the use of RNA sequencing data to quantify these general principles \citep{Mayer2022, Hoehn2019}. Identifying clonal families with high accuracy is paramount in such approaches as it avoids the potential biases of different family sizes and varying levels of clonality.

The algorithm for clonal family identification presented here is a robust inference method that enables a reliable partition of a memory B-cell repertoire into independent lineages. Using synthetic datasets we demonstrated it is distinguished by consistently high precision and high sensitivity across different junction lengths and levels of clonality. It is therefore a useful tool to explore the diversity of the repertoires and improves our ability to interpret repertoire sequencing datasets.

\section{Methods} 

\subsection{Data preprocessing and alignment} 
\label{Methods-alignment}
We focus the analysis high-throughput RNA sequencing data of IgH-coding genes \citep{Briney2019}.
The sequences were barcoded with unique molecular identifiers (UMI) to correct for the PCR amplification bias {\TM and correct sequencing errors}. 
We aligned raw sequences using presto of the Immcantation pipeline \citep{VanderHeiden2014} with tools allowing for correcting errors in UMIs and deal with insufficient UMI diversity. Reads were filtered for quality and paired using default presto parameters. We selected only sequences aligned with the IgG primer and therefore the lineage analysis is limited to the IgG subset of the repertoire. Pre-processed data was then aligned to V, D and J templates from IMGT \citep{IMGT} database using IgBlast \citep{Ye2013}. {\TM After processing, all UMI count information is discarded and only unique nucleotide sequences are kept for further analysis.}

Pairs of sequences stemming from the same VDJ recombination are expected to have the same CDR3 length $l$ and align to the same V and J {\GA templates}.  An exception could be caused by a insertion or deletion within the CDR3 that would alter its length as a result of the somatic hypermutation process. Such {\TM indel} events are rare and generally selected against \citep{lupo2021learning}, therefore in what follows we shall assume the effect of these events is negligible. 
The inference could be also affected by the misalignment of either V or J templates but we previously found the effect of alignment errors to be insignificant for identifying VJ classes \citep{spisak2020learning} (the alignment of the D template is error-prone and unreliable, hence not used in the inference procedure). 
Importantly, the two simplifications described here would result in decreased sensitivity of inference but are not expected to affect its precision.

\subsection{Modeling junctional diversity} 
\label{Methods-Ppost}
The extraordinary diversity of VDJ rearrangments can be efficiently described and quantified using probabilistic models of the recombination process as well as subsequent purifying selection. Sequence-based models can assign to each receptor sequence $s$, its total {probability of generation}, $\Pgen(s)$ \citep{Murugan2012, Elhanati2015, Marcou2018} as well as a {selection factor} $Q(s)$, inferred so as to match frequencies $\Pdata(s)$ of the sequences with a model-based distribution \citep{Elhanati2014,Sethna2020,Isacchini2021} 
\begin{equation} 
\Ppost(s) = Q(s) \Pgen(s).
\end{equation}

The $\Pgen$ model was inferred using unmutated out-of-frame sequences from \cite{Briney2019} using the IGoR software \citep{Marcou2018}. 
The selection function $Q$ model was learned using unmutated productive IgM sequences from \cite{Briney2019} using the soNNia software \citep{Isacchini2021}.

The post-selection distribution $\Ppost$ describes the diversity of the CDR3 regions and in doing so provides an expectation of pairwise distances between {unrelated}, independently generated sequences of same length $l$  \citep{Isacchini2021}. {\TM As the soNNia software does not include somatic hypermutations, the underlying assumption is that additional diversity on the CDR3 caused by hypermutations doesn't affect the distribution of pairwise distances. This assumption is justified by the quality of the fit}.
We can define
\begin{equation}
P_{\rm F}(n|l) = \left\langle \delta_{\left|s_1-s_2\right|,\;n} \right\rangle_{s_1,s_2\,\sim \Ppost(\cdot|l)},
\label{P0x}
\end{equation}
where $\left|s_1-s_2\right|$ stands for (Hamming) distance between sequences $s_1$ and $s_2$. This definition of the {null distribution} is a straightforward recipe for its estimation using (Monte Carlo) samples from $\Ppost$. 

Should $\Ppost$ differ significantly from the empirical frequencies $\Pdata$ one can resolve to the following alternative
\begin{equation}
P'_{\rm F}(n|l) = \left\langle \delta_{\left|s_1-s_2\right|,\;n} \right\rangle_{s_1\sim \Ppost(\cdot,l),\,s_2\sim \Pdata(\cdot|l)},
\end{equation}
the equivalent of the {negation distribution} as defined in \cite{lindenbaum2021alignment} and used in our evaluation of the {alignment-free method} \citep{lindenbaum2021alignment} in the method benchmark analysis.

\subsection{Estimation of pairwise prevalence} 
\label{Methods-prevalence}
Pairwise prevalence is defined as the ratio of pairs of related sequences to the total number of pairs of sequences in a given set. Related sequences share an ancestor and have diverged by independent somatic mutations, post-recombination.
Low prevalence can be a major difficulty for any inference procedure as any misassignment (or fall-out) will result in a drastic loss of sensitivity or precision. 
It is instrumental to have an {a priori} estimate of pairwise prevalence before the families are identified. 

To estimate the prevalence from the distribution of distances $P(n)$ for a given set of sequences (typically a \VJl class or $l$ class) we propose the following expectation-maximization procedure. We stipulate the distribution in question is a mixture distribution of two components, $P_{\rm F}(n)$, the expectation for unrelated sequences defined as above, and $P_{\rm T}(n)$, describing related sequences, modeled using a Poisson distribution
\begin{equation}
P_{\rm T}(n) = \frac{(\mu l)^n}{n!}e^{-\mu l},
\label{poisson}
\end{equation}
where $\mu$ is the mean divergence per basepair. {\GA If a particular CDR3 length $l$ is represented by unusually large number of \VJl classes, the} resultant shape of the positive distribution is often closer to a geometric profile, and is then modeled using $P_{\rm T}(n)=(1-M)M^n$, where $M=\frac{1}{1+\mu l}$. In sum
\begin{equation}
P(n) = \rho P_{\rm T}(n) + (1-\rho) P_{\rm F}(n).
\end{equation}
In a standard fashion, we proceed iteratively by calculating the expected value of the log-likelihood (pairs of sequences indexed by $i$)
\begin{align}
&Q(\rho,\mu | \rho_t,\mu_t) =  \nonumber\\ 
&\sum_i P_t(i\in {\rm T}) \log P_{\rm T}(n_i|\mu)+P_t(i\in {\rm F}) \log P_{\rm F}(n_i),
\end{align}
where the {membership probabilities} are defined as
\begin{align}
P_t(i\in {\rm T}) &= P(i\in {\rm T}|n_i,\mu_t,\rho_t) \\
\nonumber\\
&= \frac{\rho_t P_{\rm T}(x|\mu_t)}{\rho_t P_{\rm T}(x|\mu_t) + (1-\rho_t)P_{\rm F}(x)} \\
\nonumber\\
P_t(i\in {\rm F}) &= P(i\in {\rm F}|n_i,\mu_0,\rho_0) = 1 - P_t(i\in {\rm T}).
\end{align}
We then find the maximum
\begin{equation}
\mu_{t+1}, \rho_{t+1} = \text{argmax}\; Q(\rho,\mu | \rho_0,\mu_0) 
\end{equation}
and iterate the expectation and maximization steps until convergence, $\left|\rho_{t+1} - \rho_t\right|<\epsilon$, to obtain $\hat{\rho} = \rho_{t+1}$. 

Results for largest \VJl class within each $l$ class can be found in \SIFIG{figure2}{figure3} and  results for $l$ classes using a geometric distribution can be found in \SIFIG{figure2}{figure6}. Dependence of maximum likelihood prevalence estimates $\hat{\rho}$ on class size $N$ is plotted in \SIFIG{figure2}{figure7}.

\subsection{HILARy-CDR3} 
\label{Methods-CDR3}
The standard method for CDR3-based inference of lineages proceeds through single-linkage clustering with a fixed threshold on {\TM normalized} Hamming distance divergence {\TM (fraction of differing nucleotides)} \citep{Kepler2013, Uduman2014, Yaari2015, Nourmohammad2019}. This crude method suffers from inaccuracy as it loses precision in the case of highly mutated sequences and junctions of short length (see \SIFIG{figure4}{figure10}). 
If junctions are stored in a prefix tree data structure \citep{Knuth2013} single-linkage clustering can be performed without comparing all pairs and hence is typically orders of magnitude faster than alternatives. 
The prefix tree is a search tree constructed such that all children of a given node have a common prefix, the root of the tree corresponding to an empty string, and leaves corresponding to unique sequences to be clustered. To find neighbors of a given sequence it suffices to traverse the prefix tree from the corresponding leaf upwards and compute the Hamming distance at branchings. This method limits the number of unnecessary comparisons and greatly improves the speed of Hamming distance-based clustering \citep{boytsov2011indexing}.
We implement the prefix tree structure to accommodate CDR3 sequences. Briefly, all the CDR3 sequences of identical length are stored in the leaves of a prefix tree \citep{navarro2001guided,boytsov2011indexing}, implemented as a quaternary tree where each edge is labeled by a nucleobase (A, T, C, or G). The neighbors of a specific sequence are found by traversing the tree from top to bottom, exploring only the branches that are under a given Hamming distance from the sequence. Clusters are obtained by iterating this procedure and removing all the neighbors from the prefix tree until no sequences remain. The package is coded in C++ with a Python interface and is available independently. The time performance of this method for high-sensitivity and high-specificity partitions is studied as a part of the method  benchmark analysis.

We take advantage of the speed of a prefix tree-based clustering to perform single-linkage clustering. {\TM Besides the algorithmic speed-up afforded by the prefix tree, the difference with previous methods is that we use an adaptive threshold}. For any dataset, we define two CDR3-based partitions, high-sensitivity and high-precision clustering,  {\TM corresponding to two choices of threshold}.

{\TM The high-precision partition is obtained by setting the threshold $t$ to $t^*_{\rm prec}$ as the largest $t$ such $\hat \pi(t)\leq \pi^*$, with $\pi^*=0.99$ (99\% precision), where $\hat\pi(t)$ is given by Eqs.~\ref{precision}-\ref{sensitivity}.
 To get the high-sensitivity partition, we set the threshold to $t_{\rm sens}^*$, the smallest $t$ such that $\hat s(t)\geq s^*$, where $s^*=0.9$ (90\% sensitivity), where $\hat s(t)$ is given by Eq.~\ref{sensitivity}.

We apply these thresholds to the single linkage clustering described above to generate the precise and sensitive partitions, which are then used by the mutations-based method to find an optimal partition that merges the fine clusters within the coarse clusters (\METH{Methods-mutations} and \SIFIG{figure3}{figure8}). We refer to the high-precision partition from the CDR3 alone as HILARy-CDR3, and the mutation-based method as HILARy-full.
}

Finally, the structure of families leads to propagation of errors that lowers the precision with respect to the a priori estimate $\hat{\pi}$. Denoting family size as $z$, one error accounted for in $\FP$ causes, on average, $\langle z\rangle^2-1$ extra errors by merging two families. If the a priori precision $\hat{\pi}$ is high, we can neglect the second order effect of these two families simultaneously affected by other $\FP$ pairs). Therefore the expected precision (\ref{precision}) of the resulting partition reads
\begin{equation}
\langle\pi_{\text{post}}\rangle \simeq \frac{1}{1+(\langle z\rangle^2-1)(1-\hat\pi)}
\end{equation}
where we assumed $\hat{s}\simeq 1$. For $\hat{\pi}=99\%$ and $\langle z\rangle\simeq 2$ this formula gives $\langle\pi_{\text{post}}\rangle \simeq 97\%$.

\subsection{Synthetic data generation}
\label{Methods-synthetic}
To generate synthetic data we make use of the {\TM statistics of tree topologies of the} lineages identified in the high-sensitivity and high-precision regime of CDR3-based inference {\TM from the data} ({\TM yellow region above the black line in} \FIG{figure3}F). We denote the set of these lineages by $\mathcal{L}$. We assume that to good approximation the mutational process and the selection forces that shaped the mutational landscape in these lineages do not depend on the CDR3 length. 

To test the performance of different inference methods across CDR3 lengths, we build synthetic datasets of fixed length.

In the first step, we choose the number of families $N$. We then draw $N$ independent family sizes from the family size distribution of the form observed in healthy datasets
\begin{equation}
p(z) = \frac{z^{-\alpha}}{Z_{\alpha}}, 
\end{equation}
where $Z_{\alpha} = \sum_{z \ge 1} z^{-\alpha} = \zeta(\alpha,1)$.
In the next step, we assign a naive progenitor to each lineage by sampling from the $\Ppost$ distribution, selecting sequences with a prescribed length $l$ (\SIFIG{figure2}{figure4}). We then choose a lineage in the set of reconstructed lineages $\mathcal{L}$ at random amongst lineages of size $z$ (or, for large sizes, the lineage of the closest size smaller than $z$). 
{\TM To create a lineage with the same mutation patterns as the real data}, we then identify all unique mutations in the lineage from $\mathcal{L}$ {\TM using standard alignment and tree recontruction methods described in \cite{spisak2020learning}}, and for each mutation denote the labels of members of the lineage that carry it. For each mutation, this defines a configuration of labels, one of $2^z-1$ possible. 
We subsequently loop through observed configurations and choose new positions for all mutations to apply them to the synthetic progenitors of the ancestor, using the position- and context-dependent model of \cite{spisak2020learning}. The number of mutations assigned to a given configuration is rescaled by a factor $\frac{L+l}{L_0}$ where $L$ is the templated length of the synthetic ancestral sequence and $L_0$ is the templated length of the model lineage from $\mathcal{L}$. 

This way a synthetic lineage preserves all properties of the lineages of long CDR3s found in the data, particularly the mutational spectra (\SIFIG{figure2}{figure5}) except for the ancestral sequences and the identity of mutations. 

\subsection{HILARy-full} 
\label{Methods-mutations}

We compute the expected distributions of the CDR3 Hamming distance $n$, and the number of shared mutations $n_0$, under a uniform mutation rate assumption. In other words, we assume that the probability that a given position was mutated, given a mutation happened somewhere in a sequence of length $L$, equals $L^{-1}$ {(\TM we know this not to be true, see e.g. \cite{spisak2020learning}, but it allows for simple computations)}. It follows that the probability that a given position has not mutated once in a series of $n$ mutations is $(1-L^{-1})^n$. 

\subsubsection*{Expectation of $n_0$ under the null hypothesis}
For $n_0$ shared mutations, under the null hypothesis (we operate under the null hypothesis here since otherwise to estimate $n_0$ we would need to make assumptions about the law that governs B-cell phylogeny topologies), the likelihood reads
\begin{equation}
P_{\rm F}(n_0|n_1,n_2,L) = {L \choose n_0} p^{n_0} (1-p)^{L-n_0},
\end{equation}
where the probability that the same position independently mutated in series of $n_1$ and $n_2$ mutations is
\begin{equation}
p =  \left( 1- (1-L^{-1})^{n_1} \right) \left( 1- (1-L^{-1})^{n_2} \right).
\end{equation}
In the limit of large $L$, {\TM we have at leading order
\begin{equation}
p = \frac{n_1 n_2}{L^2},
\end{equation}
hence
\begin{align}
P_{\rm F}(n_0|n_1,n_2,L) &\simeq {L \choose n_0} \left( \frac{n_1 n_2}{L^2}\right)^{n_0} \left(1- \frac{n_1 n_2}{L^2}\right)^{L-n_0} \nonumber \\
&\simeq \frac{\left(\frac{n_1 n_2}{L}\right)^{n_0}}{n_0!} e^{-\frac{n_1 n_2}{L}},
\end{align}
where the last approximation assumes $n_1n_2\ll L^2$, which holds when mutation rates are small.
Therefore $P_{\rm F}(n_0|n_1,n_2,L)$ may be approximated by a Poisson distribution of parameter $\frac{n_1n_2}{L}$, yielding:
\begin{equation}
\langle n_0 \rangle_{\rm F} \simeq \frac{n_1n_2}{L}, \;\;\; \sigma_{\rm F}(n_0) \simeq \sqrt{\frac{n_1n_2}{L}}.
\label{scaling-n0}
\end{equation}
}

\subsubsection*{Expectation of $n$ under the hypothesis of related sequences}
The $n$ divergence of two CDR3s is interpreted as divergent mutations under the hypothesis that $s_1$ and $s_2$ are related. These mutations were harbored in parallel with $n_L = n_1 + n_2 - 2 n_0$ mutations that occurred in the templated regions ($n_0$ mutations arrived before the divergence of the two sequences began).    

Under the assumption of a uniform mutation rate, the $n_L$ mutations inform the prediction of the number of mutations expected in the CDR3. Indeed, they are related through a hidden variable, the expected number of mutations per base pair, denoted $\mu$. Integrating over this quantity we obtain
\begin{equation}
P_{\rm T}(n | n_L, l , L) = \int_0^\infty \,d\mu\, P_{\rm T}(n|\mu, l) P_{\rm T}(\mu | n_L, L),
\end{equation}
where we convolute the positive distribution (\ref{poisson}),
\begin{equation}
P_{\rm T}(n|\mu, l) = \frac{\left(\mu l \right)^{n}}{n!} e^{\mu l}
\end{equation}
and, using the Bayes rule under uniform prior over $\mu$,
\begin{equation}
P_{\rm T}(\mu| n_L, L) = L^{-1} P_{\rm T}(n_L | \mu, L) = \frac{\left(\mu L \right)^{n_L}}{n_L! L} e^{\mu L}.
\end{equation}
The result is a negative binomial distribution,
\begin{equation}
P_{\rm T}(n | n_L, l , L) = \left(\frac{L}{l+L}\right)^{n_L+1} \left(\frac{l}{l+L}\right)^{n} {n+n_L \choose n},
\end{equation}
with
{\GA
\begin{equation}
\langle n \rangle_{\rm T} = \frac{l}{L}\left(n_L+1\right), \;\;\; \sigma_{\rm T}(n) = \frac{1}{L} \sqrt{l(l+L)(n_L+1)}.
\label{scaling-n}
\end{equation}
}

\subsubsection*{Merging fine-partition clusters}

{\TM HILARy-full} relies on the results (\ref{scaling-n}) and (\ref{scaling-n0}) to define the rescaled variables (\ref{xyeq})
{\GA
\begin{equation} 
x'=\frac{n-\langle n\rangle_{\rm T}}{\sigma_{\rm T}(n)}, \quad y=\frac{n_0-\langle n_0\rangle_{\rm F}}{\sigma_{\rm F}(n_0)}.
\end{equation} 
}
{\TM
We expect $y\approx 0$, $x'>0$ for unrelated sequences, and $x'\approx 0$, $y>0$ for related sequences. So we expect $x'-y>0$ for unrelated sequences, and $x'-y<0$ for related sequences. 
{\TM We use $x'-y$ as a distance for single linkage clustering, with adaptive threshold to control performance.
The threshold $t'$ is chosen to achieve a desired precision of $\pi^*=0.99$ as in HILARy-CDR3. To this end we use soNNia-based estimate of null distribution $P_F(n|l)$ (\ref{P0x}), the data-derived distribution of the number of mutations, $P(n_1)$, and further assume $n_0 \sim \frac{n_1 n_2}{L}$ to compute the null distribution $P_F(x'-y|l)$. We can now choose a target $\pi^{*}$ and compute $t'$ such that $\hat{\pi}(t')=\pi^{*}=0.99$ using equations \ref{precision}-\ref{sensitivity}, the prevalence $\hat{\rho}$ inferred as explained earlier in the CDR3-based method, and assuming $\hat{s}\simeq 1$. As the computation of $t'$ depends on the inferred prevalence, we use this procedure only for \VJl classes with enough sequences for a reliable $\hat{\rho}$ (\SIFIG{figure3}{figure9}), namely for sizes larger than 100. For smaller sizes the treshold was set to the default value of 0.
}

To reduce the number of pairwise computations, we do not apply single-linkage clustering directly, but instead merge fine-partition clusters within coarse-partition clusters, where the fine and coarse partitions were previously obtained using the CDR3-based method (see section~\ref{Methods-CDR3}). Specifically, we compute $x'-y$ for all pairs of sequences that belong to the same coarse cluster, but to different fine clusters. Two fine-partition clusters are then merged if there exist any two sequences belonging to each of the two clusters for which {\GA$x'-y<t'$. Note that this is equivalent to performing single-linkage clustering on all sequences using the distance $-\infty$ for pairs inside a precise cluster and $x'-y$ otherwise. }
}

\subsection{Evaluation methods}
\label{Methods-evaluation}
In this section, we introduce the variation of information $v$, used for evaluating alternative methods for clonal family inference in the benchmark analysis. It is a useful summary statistic to quantify the performance of inference as it is affected by its precision as well as sensitivity \citep{brown2007automated}.
Variation of information $v(r,r^*)$ measures the information loss from the true partition $r^*$ to the inference result $r$ \citep{zurek1989thermodynamic,meilua2003comparing}. To define the variation of information we first introduce the entropy $S(r)$ of a partition $r$ of $N$ sequences into clusters $c$ as
\begin{equation}
S(r) = -\sum_{c\in r} \frac{n(c)}{N} \log{\frac{n(c)}{N}},
\end{equation}
where $n(c)$ denotes the number of sequences in cluster $c$. The mutual information between two partitions $r$ and $r^*$ can then be computed as 
\begin{equation}
I(r,r^*) = \sum_{c\in r} \sum_{c^* \in r^*} \frac{n(c,c^*)}{N} \log{\frac{n(c,c^*)}{N}},
\end{equation}
where $n(c,c^*)$ denotes the number of overlapping elements between cluster $c$ in partition $r$ and cluster $c^*$ in partition $r^*$. Finally, variation of information is given by
\begin{equation}
v(r,r^*) = S(r)+S(r^*) - 2I(r,r^*).
\end{equation}
Variation of information is a metric in the space of possible paritions since it is non-negative, $v(r,r^*)\ge0$, symmetric, $v(r,r^*)=v(r^*,r)$, and obeys the triangle inequality, $v(r_1,r_3) \le v(r_1,r_2) + v(r_2,r_3)$ for any 3 partitions \citep{zurek1989thermodynamic}.\\

\subsection{Code and data availability} 
\label{Data-availability}
{\TM We used version {\GA 1.2.0} for spectral SCOPer, 1.3.0 for SCOper using the V and J mutation presented in \SIFIG{figure4}{figure12}, version 1.2.0 for HILARy, version 0.16.0 for partis, and the code from this repository \url{https://bitbucket.org/kleinstein/projects/src/master/Lindenbaum2020/Example.ipynb} for the alignment free method.
  The HILARy tool with Python implementations of the CDR3 and mutation-based methods introduced above can be found at \url{https://github.com/statbiophys/HILARy}. The standalone prefix tree implementation can be found at \url{https://github.com/statbiophys/ATrieGC}. A complete guide to our benchmark procedure can be found in the README of the folder \url{https://github.com/statbiophys/HILARy/tree/main/data_with_scripts}, where we make available scripts to infer lineages and reproduce the benchmark figures of this article. We also upload this folder with all input and output data at {\GA \url{https://zenodo.org/records/10676371}.}}

\section*{Acknowledgements}
The study was supported by European Research Council COG 724208 and ANR-19-CE45-0018 `RESP- REP' from the Agence Nationale de la Recherche grants and DFG grant CRC 1310 `Predictability in Evolution'.
\nolinenumbers

\bibliographystyle{pnas}

\begin{thebibliography}{10}

\bibitem{Briney2019}
Briney B, Inderbitzin A, Joyce C, Burton DR
\newblock (2019) Commonality despite exceptional diversity in the baseline
  human antibody repertoire.
\newblock \emph{Nature} 566:393--397.

\bibitem{Hozumi1976}
Hozumi N, Tonegawa S
\newblock (1976) Evidence for somatic rearrangement of immunoglobulin genes
  coding for variable and constant regions.
\newblock \emph{Proc Natl Acad Sci USA} 73:3628--3632.

\bibitem{Schatz2011}
Schatz DG, Swanson PC
\newblock (2011) V({{D}}){{J}} recombination: Mechanisms of initiation.
\newblock \emph{Annual review of genetics} 45:167--202.

\bibitem{victora2022germinal}
Victora GD, Nussenzweig MC
\newblock (2022) Germinal centers.
\newblock \emph{Annual Review of Immunology} 40:413--442.

\bibitem{Feng2020}
Feng Y, Seija N, Di~Noia JM, Martin A
\newblock (2020) {{AID}} in antibody diversification: {{There}} and back again.
\newblock \emph{Trends in Immunology} pp 1--15.

\bibitem{DeBoer2001}
De~Boer RJ, Freitas A, Perelson AS
\newblock (2001) Resource competition determines selection of {{B}} cell
  repertoires.
\newblock \emph{Journal of theoretical biology} 212:333--343.

\bibitem{tas2016visualizing}
Tas JM, {et~al.}
\newblock (2016) Visualizing antibody affinity maturation in germinal centers.
\newblock \emph{Science} 351:1048--1054.

\bibitem{Mesin2016}
Mesin L, Ersching J, Victora GD
\newblock (2016) Germinal center b cell dynamics.
\newblock \emph{Immunity} 45:471--482.

\bibitem{kreer2020longitudinal}
Kreer C, {et~al.}
\newblock (2020) Longitudinal isolation of potent near-germline
  {{SARS}}-{{C}}o{{V}}-2-neutralizing antibodies from {{C}}ovid-19 patients.
\newblock \emph{Cell} 182:843--854.

\bibitem{nielsen2020human}
Nielsen SC, {et~al.}
\newblock (2020) Human {{B}} cell clonal expansion and convergent antibody
  responses to {{SARS}}-{{C}}o{{V}}-2.
\newblock \emph{Cell host \& microbe} 28:516--525.

\bibitem{Yaari2012a}
Yaari G, Uduman M, Kleinstein SH
\newblock (2012) Quantifying selection in high-throughput {{Immunoglobulin}}
  sequencing data sets.
\newblock \emph{Nucleic Acids Research} 40:e134.

\bibitem{Yaari2015}
Yaari G, Kleinstein SH
\newblock (2015) Practical guidelines for {{B}}-cell receptor repertoire
  sequencing analysis.
\newblock \emph{Genome Medicine} 7:121.

\bibitem{ortega2021modeling}
Ortega MR, Spisak N, Mora T, Walczak AM
\newblock (2021) Modeling and predicting the overlap of {{B}}-and {{T}}-cell
  receptor repertoires in healthy and {{SARS}}-{{C}}o{{V}}-2 infected
  individuals.
\newblock \emph{bioRxiv}.

\bibitem{turner2020human}
Turner JS, {et~al.}
\newblock (2020) Human germinal centres engage memory and naive {{B}} cells
  after influenza vaccination.
\newblock \emph{Nature} 586:127--132.

\bibitem{abdollahi2020automatic}
Abdollahi N, de~Septenville A, Davi F, Bernardes JS
\newblock (2020) Automatic generation of ground truth data for the evaluation
  of clonal grouping methods in b-cell populations.
\newblock \emph{bioRxiv}.

\bibitem{Elhanati2015}
Elhanati Y, {et~al.}
\newblock (2015) Inferring processes underlying {{B}}-cell repertoire
  diversity.
\newblock \emph{Philos Trans R Soc Lond, B, Biol Sci} 370:20140243.

\bibitem{briney2016clonify}
Briney B, Le K, Zhu J, Burton DR
\newblock (2016) Clonify: unseeded antibody lineage assignment from
  next-generation sequencing data.
\newblock \emph{Scientific reports} 6:1--10.

\bibitem{nouri2020somatic}
Nouri N, Kleinstein SH
\newblock (2020) Somatic hypermutation analysis for improved identification of
  {{B}} cell clonal families from next-generation sequencing data.
\newblock \emph{PLoS computational biology} 16:e1007977.

\bibitem{Isacchini2021}
Isacchini G, Walczak AM, Mora T, Nourmohammad A
\newblock (2021) Deep generative selection models of {{T}} and {{B}} cell
  receptor repertoires with sonnia.
\newblock \emph{Proceedings of the National Academy of Sciences} 118.

\bibitem{lindenbaum2021alignment}
Lindenbaum O, Nouri N, Kluger Y, Kleinstein SH
\newblock (2021) Alignment free identification of clones in {{B}} cell receptor
  repertoires.
\newblock \emph{Nucleic acids research} 49:e21--e21.

\bibitem{ralph2016LikelihoodBasedInferenceCell}
Ralph DK, Matsen FA
\newblock (2016) Likelihood-{{Based Inference}} of {{B Cell Clonal Families}}.
\newblock \emph{PLOS Computational Biology} 12:e1005086.

\bibitem{nouri2018spectral}
Nouri N, Kleinstein SH
\newblock (2018) A spectral clustering-based method for identifying clones from
  high-throughput b cell repertoire sequencing data.
\newblock \emph{Bioinformatics} 34:i341--i349.

\bibitem{ralph2022inference}
Ralph DK, Matsen~IV FA
\newblock (2022) Inference of b cell clonal families using heavy/light chain
  pairing information.
\newblock \emph{PLOS Computational Biology} 18:e1010723.

\bibitem{Horns2019}
Horns F, Vollmers C, Dekker CL, Quake SR
\newblock (2019) Signatures of selection in the human antibody repertoire:
  Selective sweeps, competing subclones, and neutral drift.
\newblock \emph{Proceedings of the National Academy of Sciences of the United
  States of America} 116:1261--1266.

\bibitem{Nourmohammad2019}
Nourmohammad A, Otwinowski J, Luksza M, Mora T, Walczak AM
\newblock (2019) Fierce {{Selection}} and {{Interference}} in {{B}}-{{Cell
  Repertoire Response}} to {{Chronic HIV}}-1.
\newblock \emph{Molecular Biology and Evolution}.

\bibitem{Saini2015}
Saini J, Hershberg U
\newblock (2015) B cell {{Variable}} genes have evolved their codon usage to
  focus the targeted patterns of somatic mutation on the complementarity
  determining regions.
\newblock \emph{Molecular Immunology} 65:157--167.

\bibitem{Mayer2022}
Mayer A, Callan CG
\newblock (2022) Measures of epitope binding degeneracy from {{T}} cell
  receptor repertoires.
\newblock \emph{bioRxiv}.

\bibitem{Hoehn2019}
Hoehn KB, {et~al.}
\newblock (2019) Repertoire-wide phylogenetic models of {{B}} cell molecular
  evolution reveal evolutionary signatures of aging and vaccination.
\newblock \emph{Proceedings of the National Academy of Sciences of the United
  States of America} 116:22664--22672.

\bibitem{VanderHeiden2014}
Vander~Heiden JA, {et~al.}
\newblock (2014) {{pRESTO}}: A toolkit for processing high-throughput
  sequencing raw reads of lymphocyte receptor repertoires.
\newblock \emph{Bioinformatics} 30:1930--1932.

\bibitem{IMGT}
Giudicelli V, {et~al.}
\newblock (2006) {{IMGT}}/{{LIGM}}-{{DB}}, the {{IMGT}}\textregistered{}
  comprehensive database of immunoglobulin and {{T}} cell receptor nucleotide
  sequences.
\newblock \emph{Nucleic Acids Research} 34:D781--D784.

\bibitem{Ye2013}
Ye J, Ma N, Madden TL, Ostell JM
\newblock (2013) {{IgBLAST}}: An immunoglobulin variable domain sequence
  analysis tool.
\newblock \emph{Nucleic Acids Research} 41:W34--W40.

\bibitem{lupo2021learning}
Lupo C, Spisak N, Walczak AM, Mora T
\newblock (2021) Learning the statistics and landscape of somatic
  mutation-induced insertions and deletions in antibodies.
\newblock \emph{arXiv preprint arXiv:2112.07953}.

\bibitem{spisak2020learning}
Spisak N, Walczak AM, Mora T
\newblock (2020) Learning the heterogeneous hypermutation landscape of
  immunoglobulins from high-throughput repertoire data.
\newblock \emph{Nucleic acids research} 48:10702--10712.

\bibitem{Murugan2012}
Murugan A, Mora T, Walczak AM, Callan CG
\newblock (2012) Statistical inference of the generation probability of
  {{T}}-cell receptors from sequence repertoires.
\newblock \emph{Proceedings of the National Academy of Sciences of the United
  States of America} 109:16161--6.

\bibitem{Marcou2018}
Marcou Q, Mora T, Walczak AM
\newblock (2018) High-throughput immune repertoire analysis with {{IGoR}}.
\newblock \emph{Nature Communications} 9:1--10.

\bibitem{Elhanati2014}
Elhanati Y, Murugan A, Callan CG, Mora T, Walczak AM
\newblock (2014) Quantifying selection in immune receptor repertoires.
\newblock \emph{Proceedings of the National Academy of Sciences}
  111:9875--9880.

\bibitem{Sethna2020}
Sethna Z, {et~al.}
\newblock (2020) Population variability in the generation and thymic selection
  of {{T}}-cell repertoires.
\newblock \emph{arXiv:2001.02843} pp 1--17.

\bibitem{Kepler2013}
Kepler TB
\newblock (2013) Reconstructing a {{B}}-cell clonal lineage. {{I}}.
  {{Statistical}} inference of unobserved ancestors.
\newblock \emph{F1000Research}.

\bibitem{Uduman2014}
Uduman M, Shlomchik MJ, Vigneault F, Church GM, Kleinstein SH
\newblock (2014) Integrating {{B}} cell lineage information into statistical
  tests for detecting selection in {{Ig}} sequences.
\newblock \emph{Journal of immunology (Baltimore, Md. : 1950)} 192:867--74.

\bibitem{Knuth2013}
Knuth DE
\newblock (2013) \emph{Art of Computer Programming, Volume 4, Fascicle 4, The:
  Generating All Trees--History of Combinatorial Generation}
\newblock (Addison-Wesley Professional).

\bibitem{boytsov2011indexing}
Boytsov L
\newblock (2011) Indexing methods for approximate dictionary searching:
  Comparative analysis.
\newblock \emph{Journal of Experimental Algorithmics (JEA)} 16:1--1.

\bibitem{navarro2001guided}
Navarro G
\newblock (2001) A guided tour to approximate string matching.
\newblock \emph{ACM computing surveys (CSUR)} 33:31--88.

\bibitem{brown2007automated}
Brown DP, Krishnamurthy N, Sj{\"o}lander K
\newblock (2007) Automated protein subfamily identification and classification.
\newblock \emph{PLoS computational biology} 3:e160.

\bibitem{zurek1989thermodynamic}
Zurek WH
\newblock (1989) Thermodynamic cost of computation, algorithmic complexity and
  the information metric.
\newblock \emph{Nature} 341:119--124.

\bibitem{meilua2003comparing}
Meil{\u{a}} M
\newblock (2003) Comparing clusterings by the variation of information.
\newblock \emph{Learning theory and kernel machines} pp 173--187.

\end{thebibliography}

\renewcommand{\thefigure}{S\arabic{figure}}
\setcounter{figure}{0}
\onecolumngrid

\clearpage
\section{Supplementary information}

\begin{figure}[H]
\centering
\includegraphics[width=0.33\linewidth]{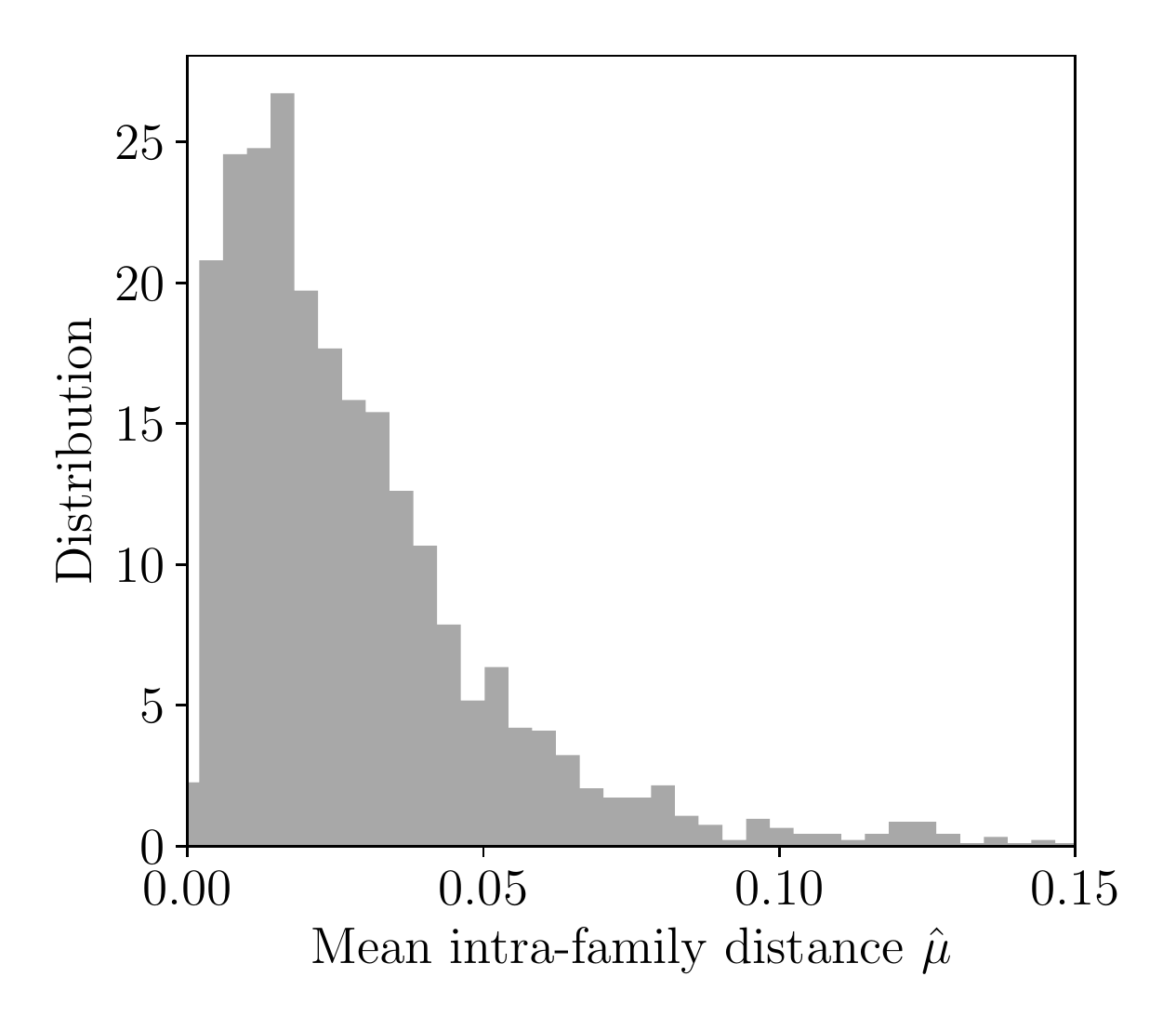}
\caption{{\bf \figtwoSonetitle} \figtwoSonecaption}
\label{figsupp:figure1}
\end{figure}

\begin{figure}[H]
\centering
\includegraphics[width=0.3\linewidth]{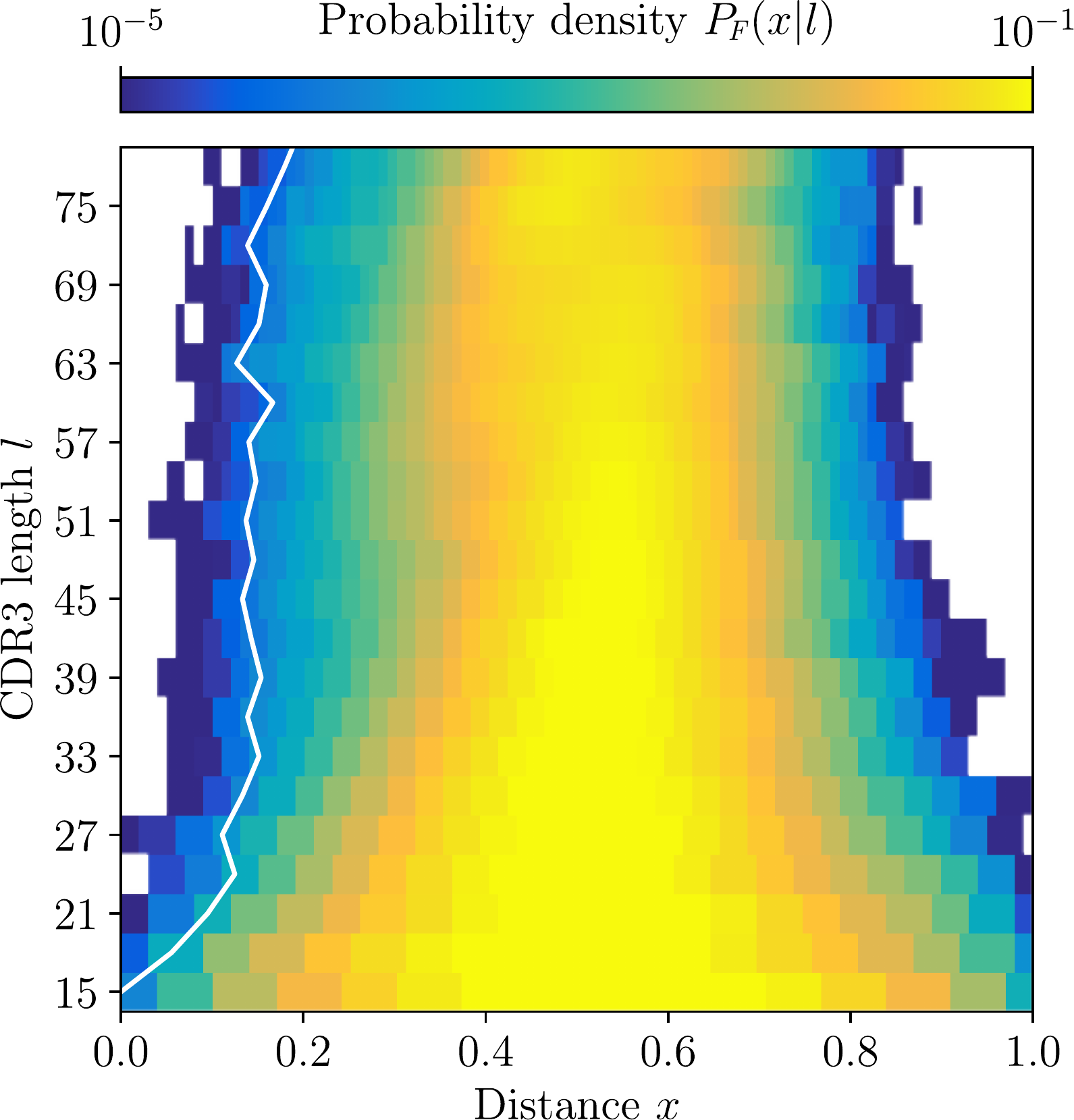}
\caption{{\bf {\figtwoStwotitle}.} \figtwoStwocaption}
\label{figsupp:figure2}
\end{figure}

\begin{figure*}
\includegraphics[width=\linewidth]{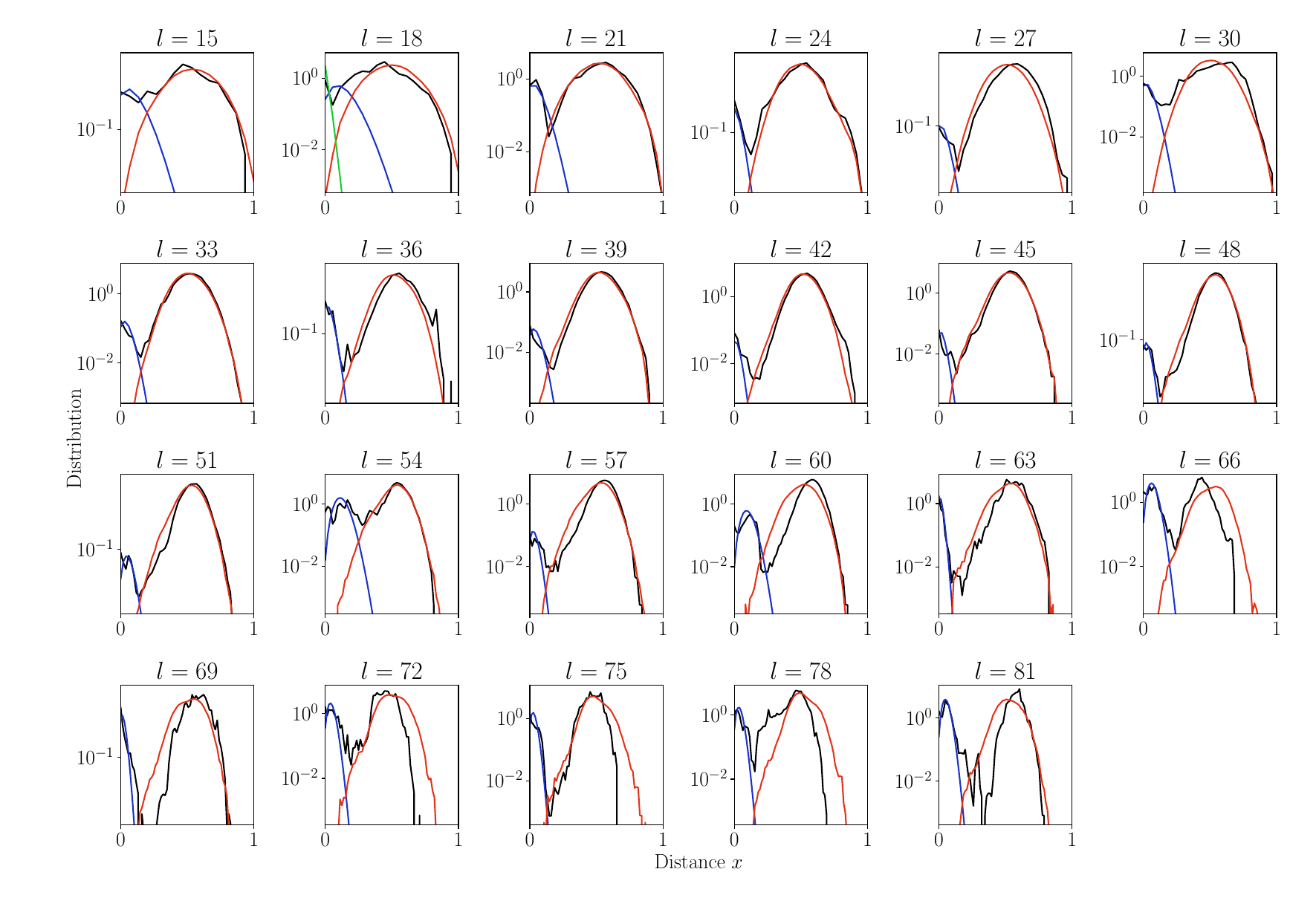}
\caption{
{\bf\figtwoSthreetitle} \figtwoSthreecaption.}
\label{figsupp:figure3}  
\end{figure*}

\begin{figure*}
\includegraphics[width=0.3\linewidth]{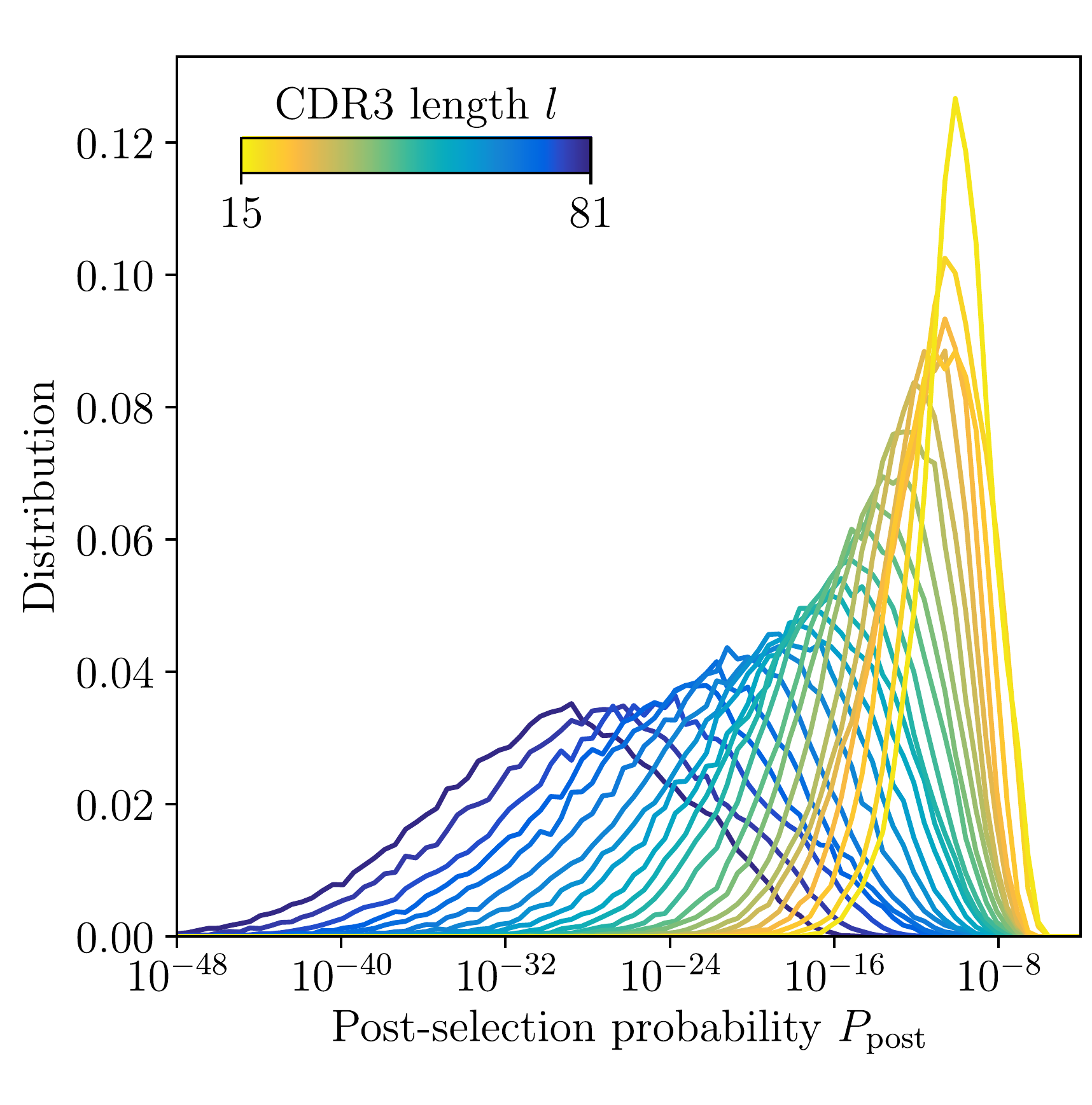}
\caption{{\bf \figtwoSfourtitle} \figtwoSfourcaption}
\label{figsupp:figure4}  
\end{figure*}

\begin{figure*}
\includegraphics[width=0.3\linewidth]{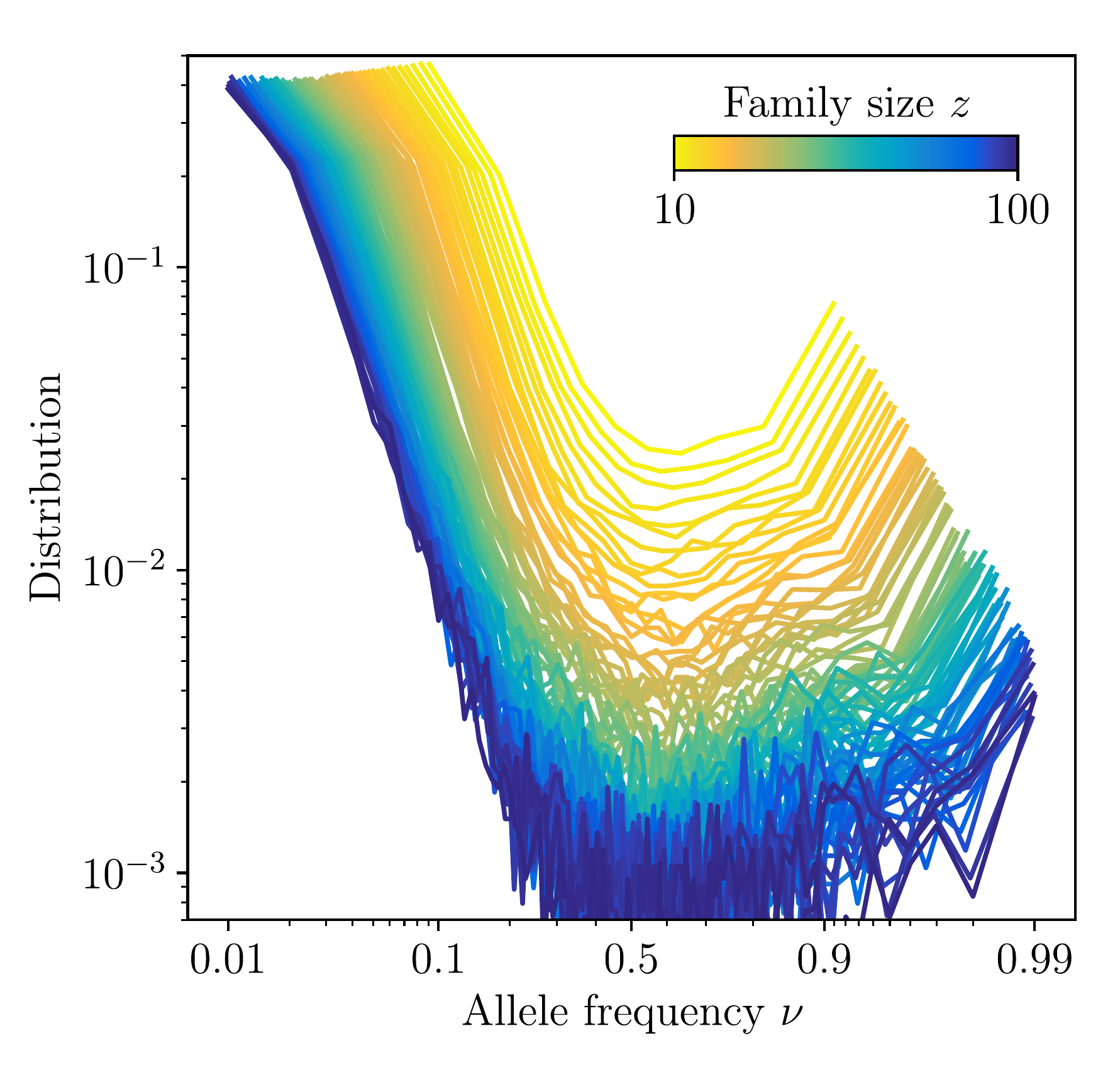}
\caption{{\bf \figtwoSfivetitle} \figtwoSfivecaption}
\label{figsupp:figure5}  
\end{figure*}

\begin{figure*}
\includegraphics[width=\linewidth]{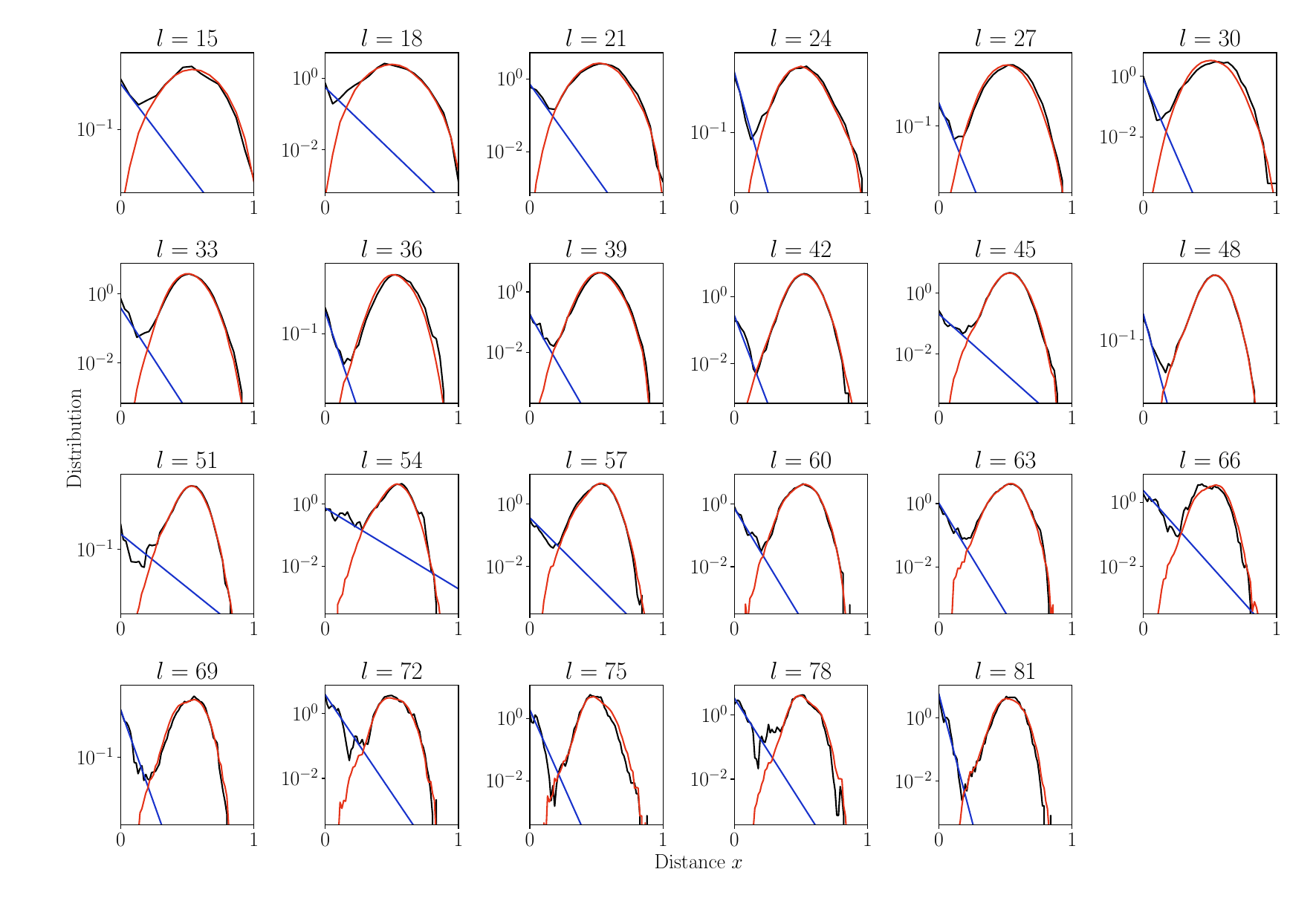}
\caption{
{\bf \figtwoSsixtitle}, \figtwoSsixcaption}
\label{figsupp:figure6}  
\end{figure*}

\begin{figure*}
\includegraphics[width=0.3\linewidth]{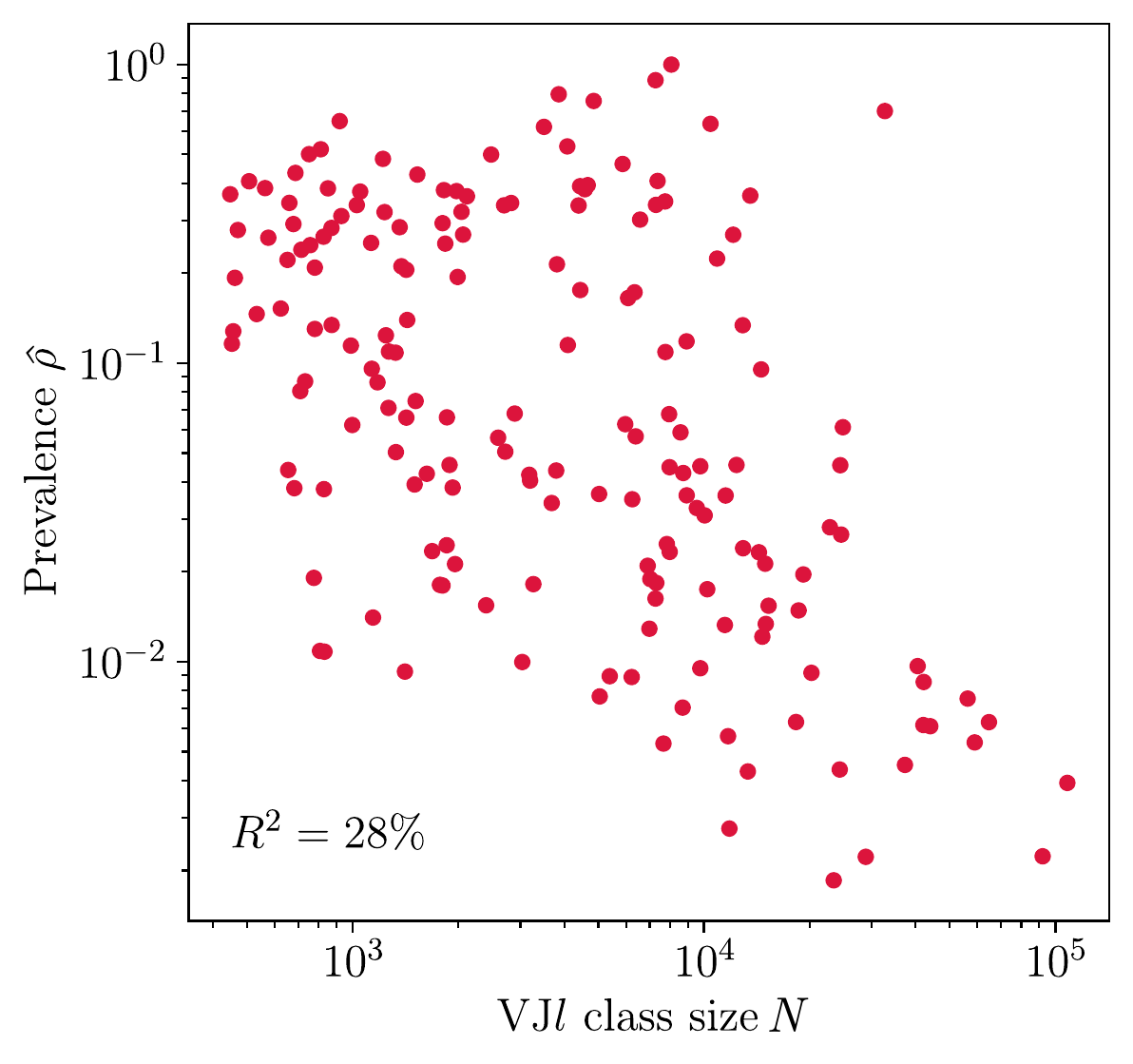}
\caption{{\bf \figtwoSseventitle} \figtwoSsevencaption}
\label{figsupp:figure7}  
\end{figure*}

\begin{figure*}
	\includegraphics[width=\linewidth]{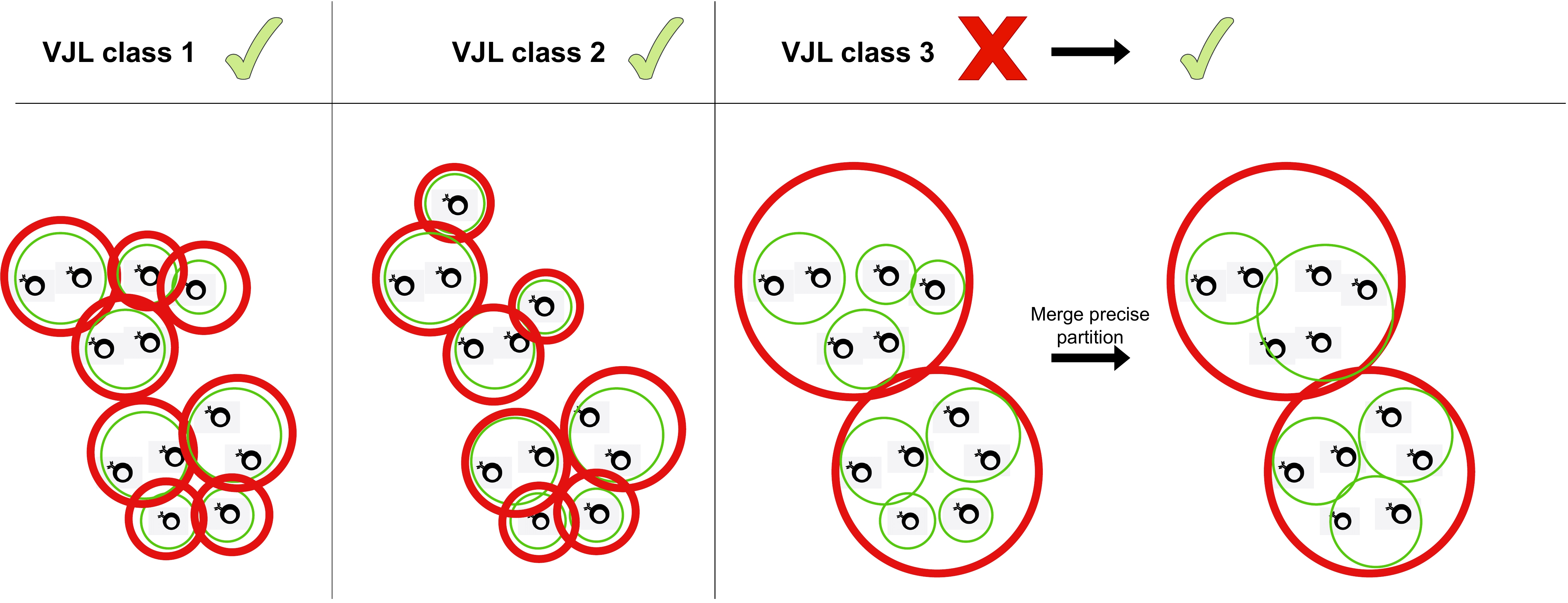}
	\caption{{\bf \figthreeSonetitle} \figthreeSonecaption}
\label{figsupp:figure8}  
\end{figure*}

\begin{figure*}
	\includegraphics[width=\linewidth]{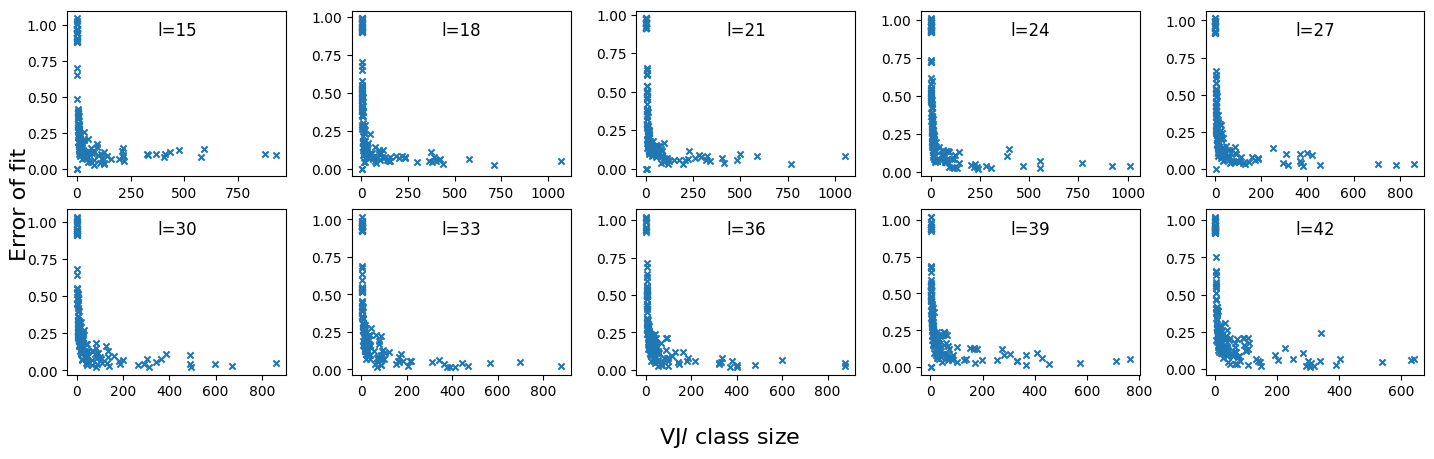}
	\caption{{\bf \figthreeStwotitle} \figthreeStwocaption}
\label{figsupp:figure9}  
\end{figure*}

\begin{figure*}
	\includegraphics[width=\linewidth]{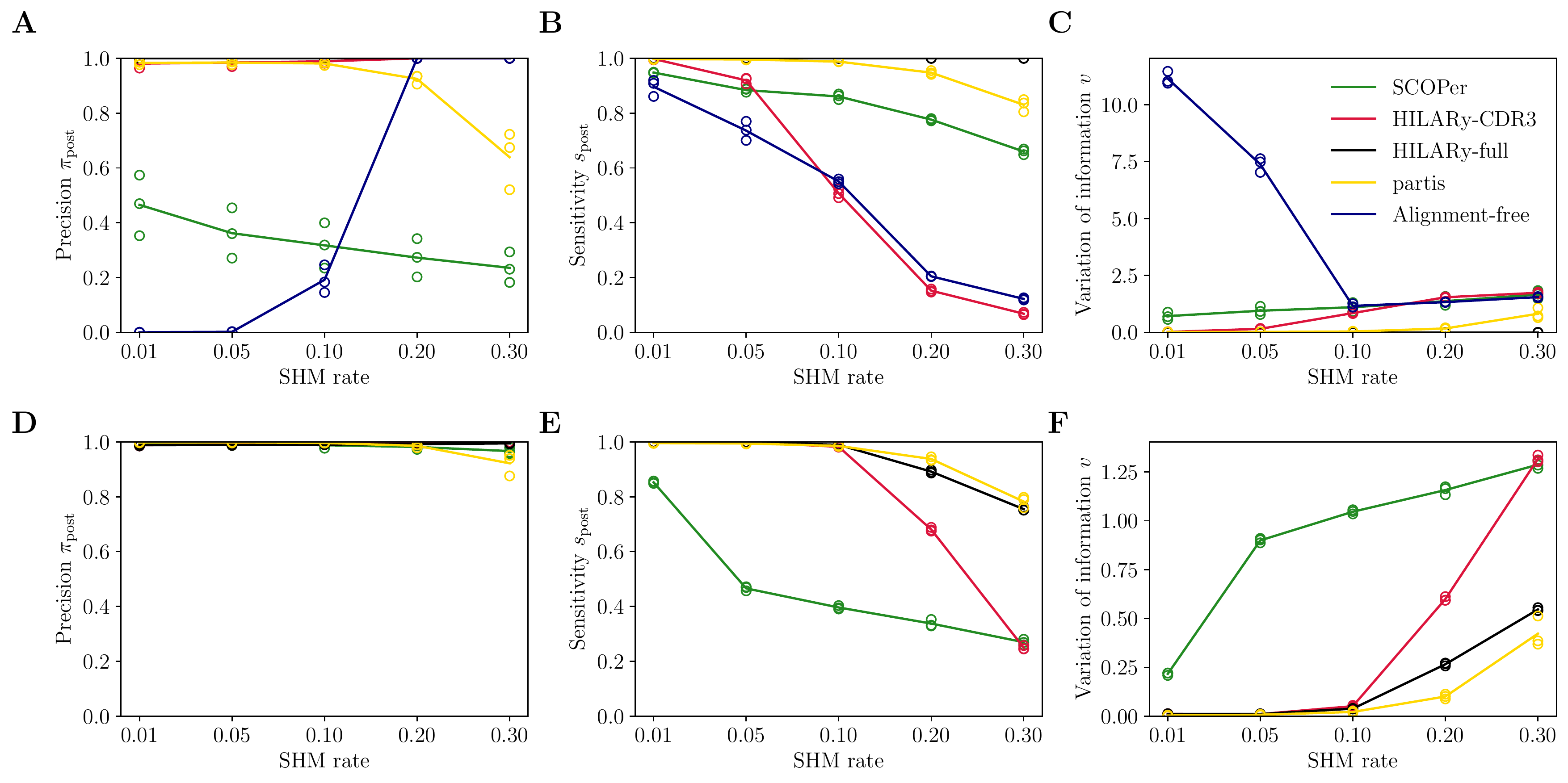}
	\caption{{\bf \figfourSonetitle} \figfourSonecaption}
\label{figsupp:figure11}  
\end{figure*}

\begin{figure*}
	\includegraphics[width=\linewidth]{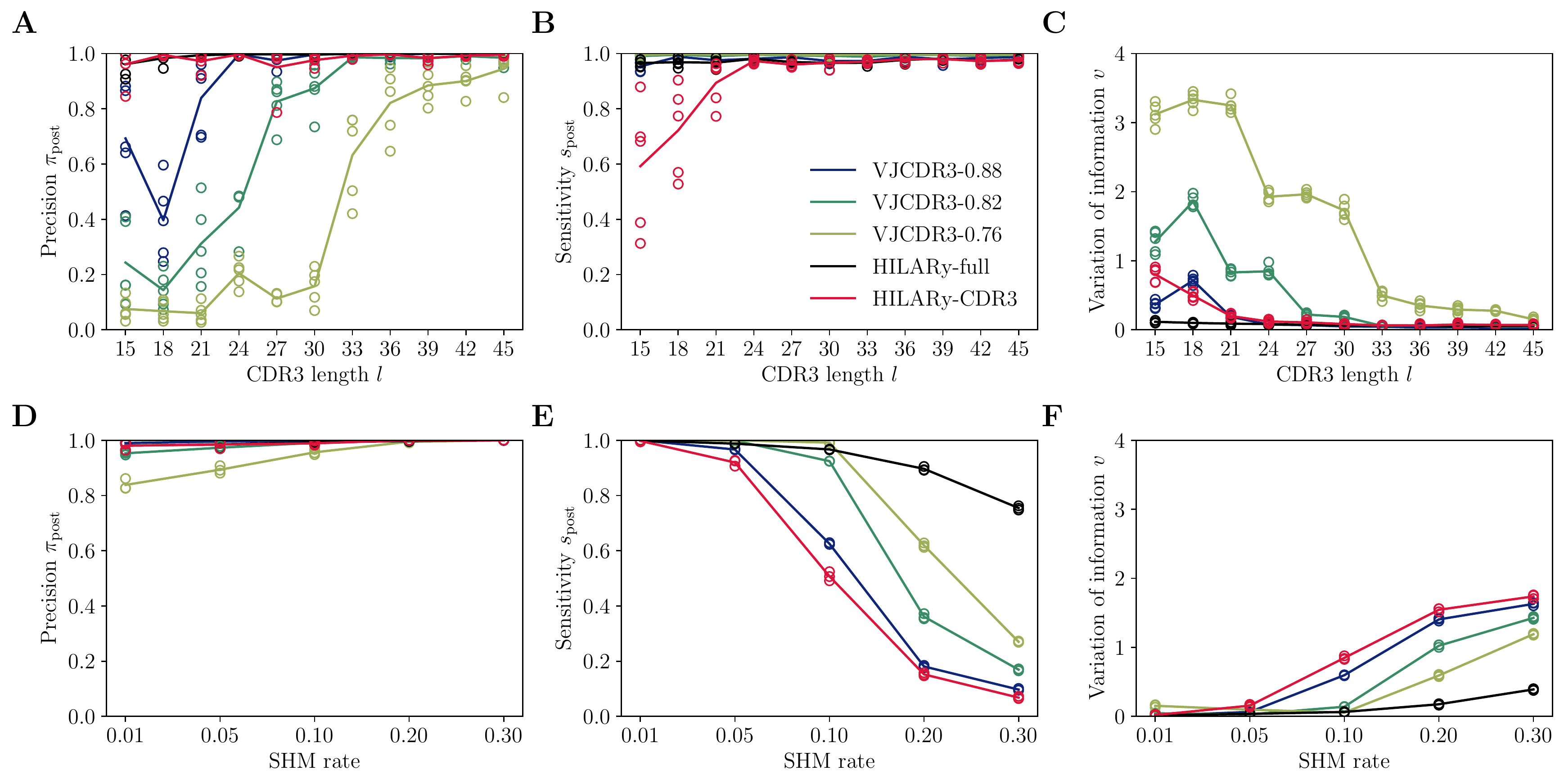}
	\caption{{\bf \figfourStwotitle} \figfourStwocaption}
\label{figsupp:figure10}  
\end{figure*}

\begin{figure*}
	\includegraphics[width=\linewidth]{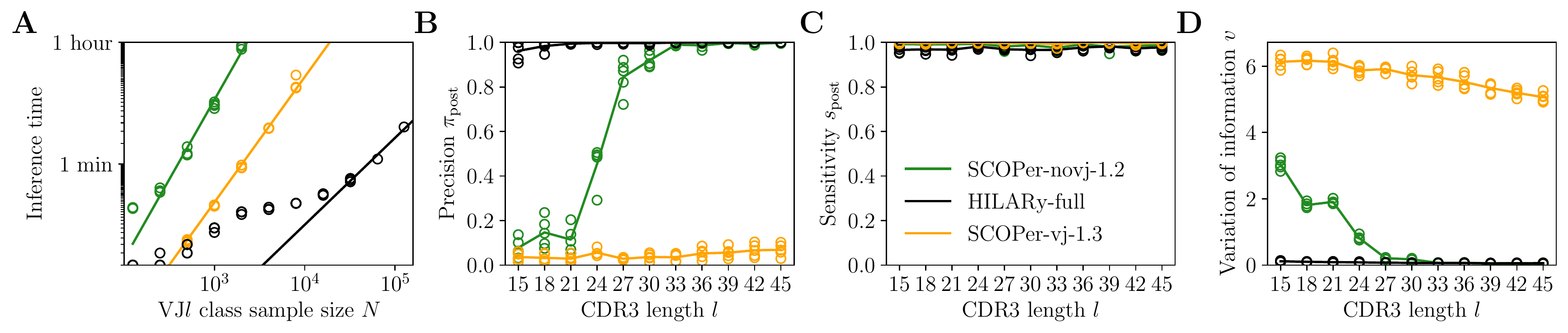}
	\caption{{\bf \figfourSthreetitle} \figfourSthreecaption}
\label{figsupp:figure12}  
\end{figure*}

\end{document}